\documentclass[a4paper,twoside,12pt]{article}

\usepackage{times}
\usepackage{amssymb,amsthm,amsmath}
\usepackage{psfrag,epsfig}
\usepackage[T1]{fontenc}
\usepackage{fancyhdr}
\usepackage{cite}
\usepackage[flushmargin]{footmisc}       

\newcommand{\matrice}[2][cccccccccccccccccc]{\left(\hspace{-2mm}\begin{array}{#1}#2\\ \end{array}\hspace{-2mm}\right)}

\renewcommand{\theequation}{\thesection.\arabic{equation}}

\let\theta=\vartheta
\let\phi=\varphi
\let\rho=\varrho
\let\epsilon=\varepsilon

\def\ie{i.e.~}
\def\eg{e.g.~}

\def\L{{\rm L}}
\def\R{{\rm R}}
\def\B{{\rm B}}
\def\F{{\rm F}}

\def\0{{}}
\def\eref#1{(\ref{#1})}
\def\Section{{Section}}

\def\Subsection{{Section}}

\def\Remark{{Remark}}

\def\Definition{{Definition}}

\def\Fig{{Figure}}
\def\Figs{{Figures}}

\newtheorem{theorem}{Theorem}[section]

\newtheorem{definition}[theorem]{Definition}
\newtheorem{remark}[theorem]{\it Remark}
\newtheorem{assumption}{Assumption A\hspace{-1mm}}

\psfragscanon

\title{{\vspace{-5mm}\bf ThermoElectric Transport Properties\\ of  a Chain of Quantum Dots\\ with Self-Consistent Reservoirs}\vspace{5mm}}

\author{Philippe A. Jacquet \vspace{5mm}\\  \small{D\'{e}partement de Physique Th\'{e}orique}\vspace{-1mm}\\ \small{Universit\'{e} de
  Gen\`{e}ve}\vspace{-1mm}\\ \small{CH-1211 Gen\`{e}ve 4, Switzerland}}

\date{}

\begin{document}

\thispagestyle{plain}

\maketitle

\thispagestyle{empty}

\begin{abstract}
\noindent
We introduce a model for charge and heat transport
based on the Landauer-B\"uttiker scattering approach. The system consists of a
chain of $N$ quantum dots, each of them being coupled to a particle
reservoir. Additionally, the left and right ends of the chain are coupled
to two particle reservoirs. All these reservoirs are independent and can be described by any
of the standard physical distributions: Maxwell-Boltzmann,
Fermi-Dirac and Bose-Einstein. In the linear response regime, and
under some assumptions, we first describe the
general transport properties of the system. Then we impose the
self-consistency condition, \ie we fix the boundary values
$(T_\L,\mu_\L)$ and $(T_\R,\mu_\R)$, and adjust the parameters $(T_i,\mu_i)$, for $i = 1,\dots,N$, so that the net average electric and heat currents into all the intermediate reservoirs
vanish. This condition leads to expressions for the temperature and chemical
potential profiles along the system, which turn out to be
independent of the
distribution describing the reservoirs. We also determine the average electric and heat currents flowing through the
system and present some numerical results, using random matrix theory,
showing that these currents are typically governed by Ohm and Fourier laws. \vspace{3mm} \vspace{3mm}
\end{abstract}

\setcounter{equation}{0}
\section{Introduction}

The study of transport properties of systems out of equilibrium is
a fascinating subject in theoretical
physics. In particular, various models have been developed to find out
what are the underlying microscopic mechanisms giving rise to
macroscopic behaviours such as Ohm and Fourier laws \cite{Jackson,BLR,LLP}. 

\begin{figure}[h!]
\begin{center}
\psfig{file=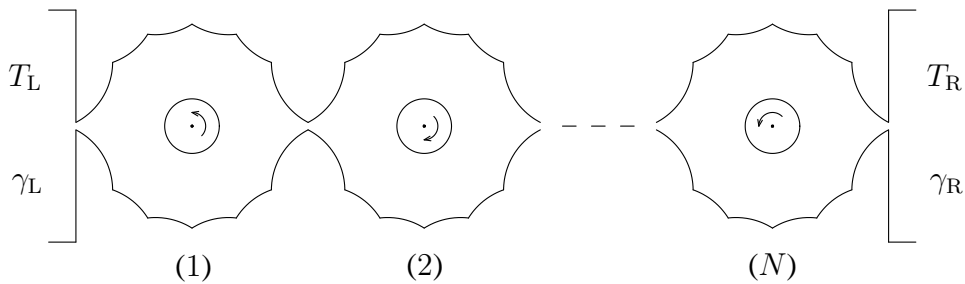,width=12cm,angle=180}
\vspace{5mm}\caption{The classical EY-model composed of $N$ cells.}
\label{The Classical System}
\end{center}
\end{figure}

One interesting
class of models that has been introduced with this purpose is
related to the Lorentz gas \cite{Lorentz,LS,LebSpohn,Wagner,KlagesPRL,Rateitschak,Carlos,LLM,EY}. In particular, a classical
model for particle and energy transport has recently been
investigated by Eckmann and Young \cite{EY}. In this EY-model, the system is a linear chain of \emph{chaotic}
cells, each one containing a fixed freely rotating disc at its centre
(see \Fig~\ref{The Classical System}). Both ends of the chain are coupled to
classical particle reservoirs assumed to be in thermal equilibrium at
temperatures $T_\L$ and $T_\R$, respectively. Particles are injected into the system from these
reservoirs at characteristic rates $\gamma_\L$ and $\gamma_\R$, respectively. Once in the system, the non-interacting
particles move freely and collide elastically with the boundaries of
the cells as well as with the discs. The particles can also leave the
system through these reservoirs. 

Note that in this model the discs play the
role of energy tanks, allowing the redistribution of
energy among the particles in a given cell and thus permitting the
system to reach at large times a stationary state satisfying local
thermal equilibrium (LTE). As a consequence, this
model admits a well-defined notion of local temperature.

One of the main results states
that, under some assumptions, the average stationary particle and energy currents through
the system are governed by Fick and Fourier laws.

In this paper, we will always assume that the considered system has reached its (unique) \emph{stationary} state and discuss only its \emph{average} transport properties. 
In particular, we shall only consider average stationary currents and consequently simply call them currents.

In the present work, we shall construct a similar type of model using
the Landauer-B\"uttiker scattering approach
\cite{Landauer,Landauer1970,ButtikerFourT,Buttiker-Scattering,ButtikerScattering}. This will
permit, in particular, to establish an effective \emph{quantum}
version of the EY-model.

Basically, in the scattering approach to charge and heat transport,
one expresses the (average) electric and heat currents through a conductor in
terms of the scattering data of this conductor. Therefore, starting from
the EY-model, we shall make the following crucial
modifications:
\begin{itemize}
\item[(1)] The main modification concerns the scattering version of
  the disc. There are two features of the EY-model that should be pointed out: (i) The
  exact position of the disc within a given cell is not important, so one
  may consider the cell represented in \Fig~\ref{The two Systems}
  (left). (ii) Once the system has
  reached a stationary state satisfying LTE, the boxes containing the discs are at
  equilibrium and consequently the net (average) particle and energy
  currents into these boxes vanish \cite{Wagner,KlagesPRL,Rateitschak}.

\begin{figure}[h!]
\begin{center}
\centerline{\psfig{file=Fig2a.ps,width=4cm}  \hspace{20mm}\psfig{file=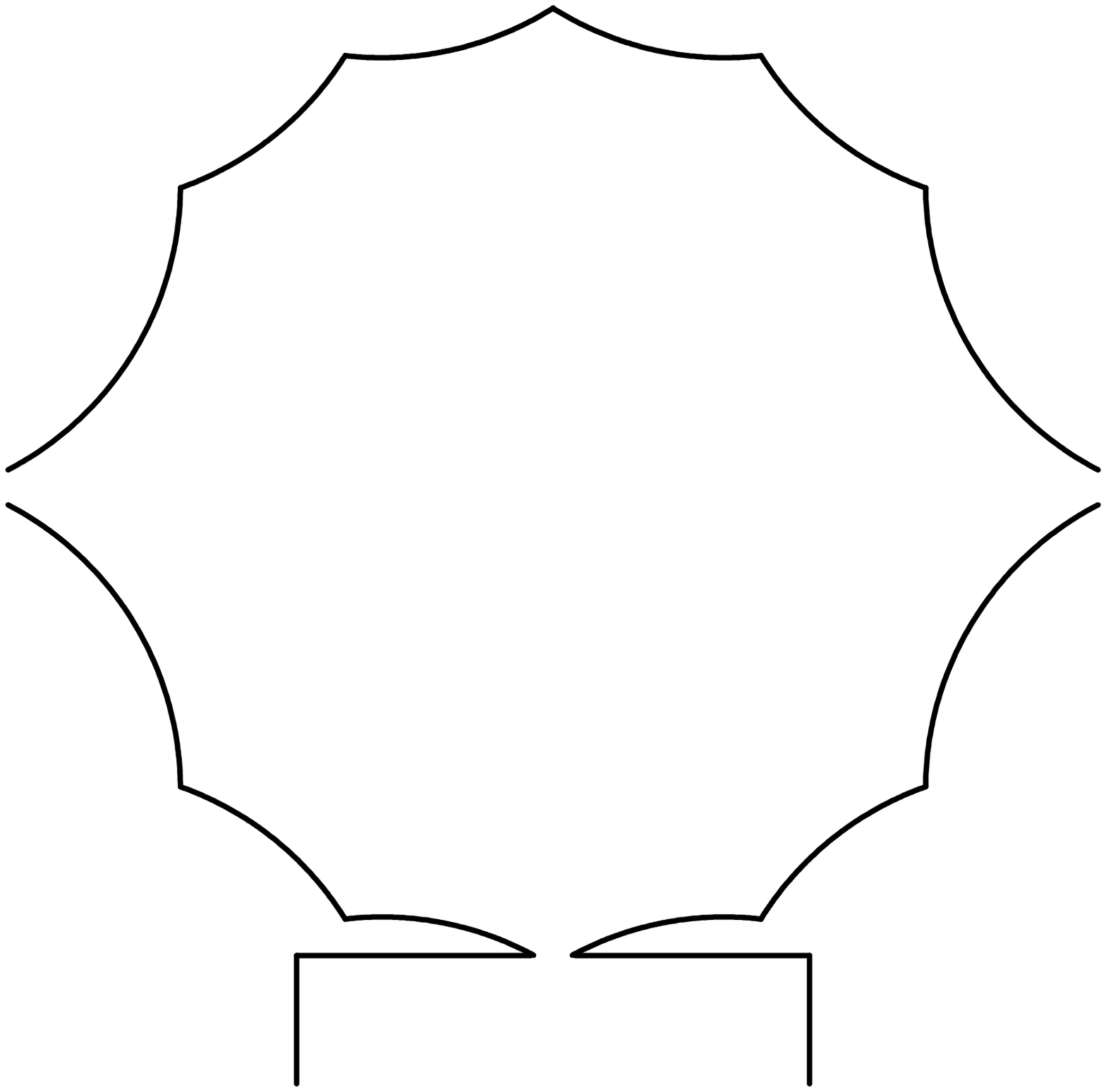,width=4cm}}
\vspace{2mm}\caption{Left: A possible EY-cell. Right: The
  corresponding scattering cell.}
\label{The two Systems}
\end{center}\vspace{-7mm}
\end{figure}

Therefore, in the scattering approach, one can obtain
  an \emph{effective} description of the discs as
  follows\footnote{This way of modelling the discs in the scattering
  approach was suggested to me by M. B\"uttiker.}: We replace the boxes containing the discs by independent particle reservoirs with characteristic temperatures and chemical potentials
  such that the net (average) electric and heat currents
  into these particle reservoirs vanish (\Figs~\ref{The two Systems} and \ref{The Quantum
  System}). In this manner, the particles in a given cell will be in
  equilibrium with their respective reservoir, \ie the system
  will be in local thermal equilibrium. 

Although the boxes containing the discs have finite energy,  while the
  reservoirs have infinite energy, we believe that the
  transport properties of the system \emph{in the stationary state} (assuming
  there is one) will be similar in both cases. 

Usually, the condition of
  zero net (average) particle and heat currents, between the system and a reservoir,
  is referred to as the
  self-consistency condition
  \cite{BolRV,RichViss,Visscher,Davies,BLebL,roy}. If one is only
  interested in the electric current, the self-consistent reservoirs correspond to
  the \emph{voltage probes} used in mesoscopic physics, originally
  introduced to model the inelastic scattering occurring in a conductor \cite{Buttiker-Reservoir,ButtikerQuantum,ButtikerBarriers,Amato,ButtikerQuantum2,Buttiker-Review,Ando,Pilgram,Heidi}. 
  
  A \emph{crucial} distinction appears here: In voltage probes \emph{only} the electric currents are set to zero while in this paper the self-consistent reservoirs require to set \emph{both} the electric and heat currents to zero. To our knowledge, this fundamental difference has never been discussed in the formalism of Landauer-B\"uttiker and consequently makes one of the novel aspects of the present work. In particular, our results generalise works by, for example, B\"uttiker \cite{Buttiker-Reservoir,ButtikerQuantum,ButtikerBarriers}, D'Amato and Pastawski \cite{Amato}, Ando \cite{Ando} and very recently Roy and Dhar \cite{roy}. 

If the couplings between the system and the self-consistent reservoirs
are sufficiently small, then one may interpret the self-consistent reservoirs as
ideal potentiometers and thermometers \cite{Engquist,Sivan-Imry}.
\item[(2)] The transport properties of the cells will be given
  in terms of $N$ scattering matrices $S^{(1)},\dots
  ,S^{(N)}$. However, we will work mostly in terms of the scattering matrix $S$ associated
  to the global multi-terminal system, which is obtained by composing the
  $N$ local scattering matrices together.
\item[(3)] In order to have some generality, we will describe the
  particle reservoirs in terms of the three
  standard physical distributions: the classical
  Maxwell-Boltzmann distribution and the quantum Fermi-Dirac
  and Bose-Einstein distributions. We assume that the
  particles may carry some charge $e$, so that we may speak of electric
  currents instead of particle currents. Nevertheless, we
  shall assume, as in the EY-model, that the particles do not interact with each
  other.
\end{itemize}

\begin{figure}[h!]
\begin{center}
\psfig{file=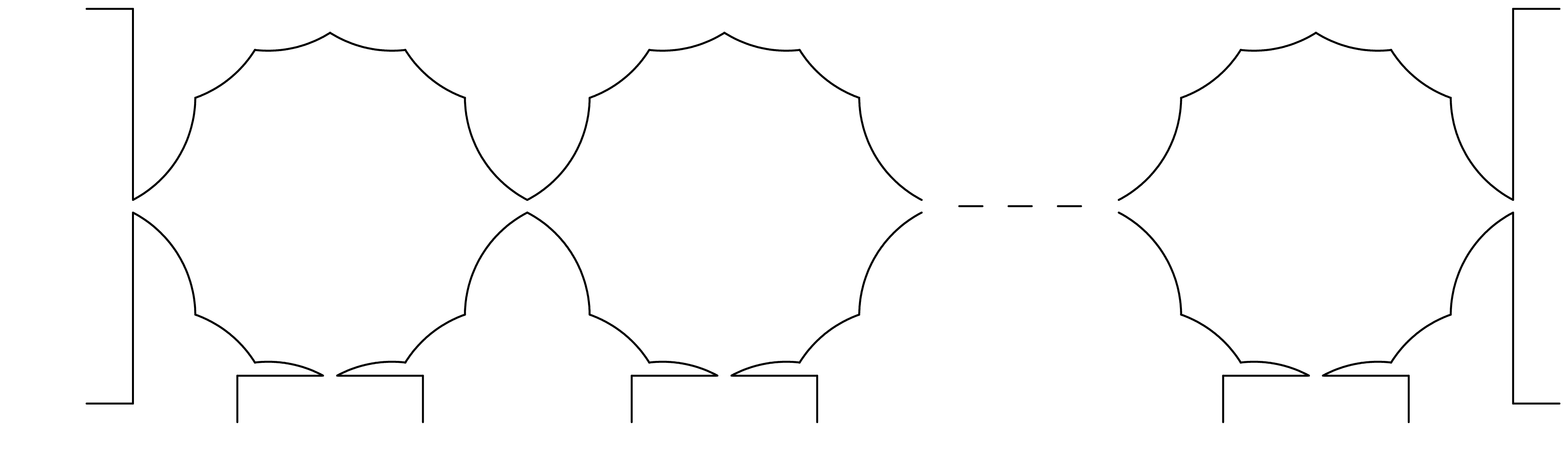,width=12cm}
\caption{The system made of $N$ quantum dots with self-consistent reservoirs.}
\label{The Quantum System}
\end{center}\vspace{-2mm}
\end{figure}

For obvious reasons, we now call the cells {\it quantum dots}.
In the EY-model, the outer boundaries of the cells were chosen
such that the
system was chaotic. We shall at some point impose this property by assuming that the quantum dots are classically
chaotic.

The organisation of this paper is as follows. In \Section~\ref{The
  model}, we establish the framework of multi-terminal systems in
  mesoscopic physics, present our model and outline the strategy
  adopted to obtain the desired chain of quantum dots. Starting with \Section~\ref{General properties}, we will work in the linear response
  regime and under the following main assumption: the scattering
  matrix $S$ associated to the system does not depend on the
  energy. As we shall see, this assumption, which is a good approximation in
  certain limiting cases, will lead to some interesting consequences. 

More precisely, the main results of this paper are divided into three parts. In \Section~\ref{General
  properties}, we present the general transport properties of the
  system, and discuss in particular the Onsager relations and the
  entropy production. In
  \Section~\ref{The null currents condition}, we solve the
  self-consistency condition, \ie we fix the boundary values
$(T_\L,\mu_\L)$ and $(T_\R,\mu_\R)$, and adjust the parameters $(T_i,\mu_i)$, for $i = 1,\dots,N$, so that the net average 
electric and heat currents into all the intermediate reservoirs
vanish. This condition leads to expressions for the temperature and chemical potential
profiles along the system, which turn out to be
independent of the nature of the particles, \ie independent of the
distribution describing the reservoirs. From these profiles, we work
out the average electric and heat currents flowing through the system. Let us emphasise that all these results are obtained
rigorously and actually hold for \emph{any}
  geometry (see \Fig~\ref{The General
    System}). These results constitute the main analytical results of this paper.

In \Section~\ref{Numerical analysis}, we restrict our attention to the
linear geometry represented in \Fig~\ref{The Quantum
  System}. In analogy with the
EY-model, we assume that the quantum
dots are classically chaotic and consequently use random matrix
theory (RMT) to describe the transport properties of the quantum
dots. While we cannot prove the validity of Ohm and Fourier laws in this
situation, some approximate derivations are feasible in some particular cases
(see \Remark~\ref{Remark Amato}). We thus turn to numerical
simulations and found that Ohm and Fourier laws typically hold (in the
sense of RMT) in a chain of
quantum dots with self-consistent reservoirs.


\setcounter{equation}{0}
\section{The Model}\label{The model}

We consider a chain of $N$ connected quantum dots, where each dot is coupled to a
particle reservoir at  temperature $T_i$ and chemical potential
$\mu_i$, with $i = 1, \dots, N$. Additionally, the left and right ends
of the chain are coupled to particle reservoirs with
parameters ($T_\L$,$\mu_\L$) and ($T_\R$,$\mu_\R$), respectively (\Fig~\ref{The Quantum
  System}). All these reservoirs, which we also call \emph{terminals},
are independent and inject particles into the
system according to some distribution function. We assume that they can also absorb
particles without changing their state. In this paper, we shall
consider the following three cases ($i \in \{\L, \R, 1, \dots, N\}$):
\begin{equation}\label{Maxwell-Boltzmann}
f^{\rm MB}_i(E) \equiv f^{\rm MB}(E; T_i, \mu_i) = \exp\left(-\frac{E-\mu_i}{k_\B
    T_i}\right)~,
\end{equation}
\begin{equation}\label{Fermi-Dirac}
f^{\rm FD}_i(E) \equiv f^{\rm FD}(E; T_i, \mu_i) = \left[\exp\left(\frac{E-\mu_i}{k_\B
    T_i}\right) + 1\right]^{-1}~,
\end{equation}

\begin{equation}\label{Bose-Einstein}
f^{\rm BE}_i(E) \equiv f^{\rm BE}(E; T_i, \mu_i) = \left[\exp\left(\frac{E-\mu_i}{k_\B
    T_i}\right) - 1\right]^{-1}~.
\end{equation}
Here, $E \in [0,\infty)$ is the energy, $k_\B$ is Boltzmann's
  constant and $T_i > 0$ is the temperature of the $i$-th terminal. For the
  chemical potentials there is one subtlety: for the Maxwell-Boltzmann and Fermi-Dirac
  functions, one has $\mu_i \in (-\infty, \infty)$, while for the Bose-Einstein
  function, one has $\mu_i \in (-\infty, 0)$. Let us emphasise that, for each $i$, the parameters $T_i$ and $\mu_i$ are
  \emph{independent} of each other.

We consider that the transport properties of the $k$-th quantum
  dot are described by a scattering matrix $S^{(k)}(E)$ at energy $E > 0$:
\begin{equation}
 S^{(k)}(E) = \left(S^{(k)}_{ij;mn}(E)\right)~, \hspace{5mm} k \in \{1, \dots, N\}~,
\end{equation}
where $i,j$ denote the three possible entrances of the $k$-th
dot and the indices $m,n$ denote their
corresponding channels. In \Section~\ref{Numerical analysis}, we
shall use these scattering matrices to build the scattering matrix
$S$ associated to the \emph{global} system made of $N$ quantum dots. For the present,
let us assume that we are given the scattering matrix of the global system:
\begin{equation}
 S(E) = \left(S_{ij;mn}(E)\right)~,
\end{equation}
where $i,j \in \{\L, \R, 1, \dots, N\}$ and for each couple $(i,j)$
the indices $m \in \{1, \dots, M_i\}$ and $n \in \{1, \dots, M_j\}$
number the channels in terminal $i$ and $j$, respectively. Therefore, the
matrix element $S_{ij;mn}(E)$ is the probability \emph{amplitude} that a particle
with energy $E$ incident in channel $n$ in terminal $j$ is transmitted
into channel $m$ in terminal $i$. 

Being given $S(E)$, one can define the \emph{total}
transmission probability $t_{ij}(E)$ that a particle with energy $E$ goes from terminal $j$ to terminal $i$ \cite{Buttiker-Symmetry}:
\begin{equation}\label{transmission prob}
t_{ij}(E) = \sum_{m=1}^{M_i}
\sum_{n=1}^{M_j} |S_{ij;mn}(E)|^2~.
\end{equation}

Throughout this paper, we always assume that the following holds.
\begin{assumption}
For each $E > 0$, the complex matrix $S(E)$ is
\emph{unitary}.\label{Assumption S}
\end{assumption}
 On the other hand, we do \emph{not} assume the scattering matrix $S(E)$ to be
  symmetric. Using the unitarity of $S(E)$ one easily obtains
the following results:
\begin{equation}
\sum_{i} t_{ij}(E) = M_j~, \hspace{2mm} \forall j \hspace{7mm} \mbox{and} \hspace{7mm} \sum_{j} t_{ij}(E) =
M_i~, \hspace{2mm} \forall i~.\label{Property Transmission}
\end{equation}

\begin{remark}
\textnormal{The sums appearing in \eref{Property Transmission} are
over the set $\{\L, \R, 1, \dots, N\}$ of \emph{all} terminals. Unless stated, in what follows every
sum  will be understood over this set.}
\end{remark}

\subsection{The Currents}

In this paper, we shall always assume that the following holds.
\begin{assumption}
\begin{itemize}
\item[(i)] The particles do not interact with each other.
\item[(ii)] The system admits a unique stationary state (out of equilibrium). 
\end{itemize}\label{Assumptions}
\end{assumption}

Under these assumptions, one can derive for any of the distributions
\eref{Maxwell-Boltzmann}--\eref{Bose-Einstein} the expressions for the
(average)  \emph{electric currents} in a
multi-terminal system \cite{ButtikerScattering,Buttiker-Review,Datta,CA}. One finds
\begin{equation}
I_i = \frac{e}{h} \sum_{j} \int_{0}^{\infty} \left[t_{ji}(E) f_i(E) - t_{ij}(E) f_j(E)\right] dE~,\label{Electrical Current 1}
\end{equation}
where $e > 0$ is the charge carried by the particles and $h$ is Planck's
constant. 

\begin{remark}
\textnormal{To our knowledge, the expression \eref{Electrical Current 1} for the average electric current in a multi-terminal system was first introduced, through formal arguments, by B\"uttiker in \cite{ButtikerScattering} and was very recently given a rigorous and general derivation (together with \eref{Current Heat}) in a work by Aschbacher \emph{et al.}~\cite{CA} using a $C^{*}$ algebraic approach.}
\end{remark}

The expression \eref{Electrical Current 1} can be interpreted as follows: $t_{ji}(E) f_i(E)$ is the average number of particles with energy $E$ that are transmitted from
terminal $i$ to terminal $j$, and $t_{ij}(E) f_j(E)$ is the same but from terminal $j$ to terminal $i$. Therefore, $I_i$
is the \emph{net} average electric current at terminal $i$, counted
positively from the terminal to the system.

\begin{remark}
\textnormal{To describe the transport of neutral particles
(such as neutrons or phonons), it suffices to set $e = 1$ in \eref{Electrical Current 1} and one will obtain the
particle current.}
\end{remark}

It turns out that most relations that we shall encounter
will  have the same form in the three considered cases (an example
being given by the
expression \eref{Electrical Current 1}). Therefore, from now on, all relations not wearing the superscript MB,
  FD or BE will be assumed to hold in all three cases.
  
We shall also 
  obtain some expressions which do not depend at all on the distribution
  function describing the reservoirs. In order to emphasise the fact that such
  relations do not depend on the nature of the particles, we shall say
  that they are \emph{universal}. We hope the reader will not be confused with the notion
  of universality used to refer to properties that are model
  independent.

Under the Assumption A\ref{Assumptions}, one can also derive the expression for
  the (average) \emph{heat current} at terminal
  $i$ \cite{Butcher,CA}. Basically, the idea is to write the particle current, which is given by \eref{Electrical Current 1} with
  $e=1$, as well as the energy current, which is also given by \eref{Electrical Current 1} with
  $e=1$ but with an extra $E$ in the integrand, and then to invoke the first law of thermodynamics $\delta Q_i = dE_i - \mu_i
  dN_i$. One finds
\begin{equation}\label{Current Heat}
J_i = \frac{1}{h} \sum_{j} \int_{0}^{\infty} \left[t_{ji}(E) f_i(E) - t_{ij}(E) f_j(E)\right] (E-\mu_i) dE~.
\end{equation}

\begin{remark}
\textnormal{As we shall see in the linear
  response analysis, if one considers $\delta \mu_i / e$ and $\delta T_i / T$ as the thermodynamic forces,
  then $I_i$ and $J_i$ turn out to be the right currents to obtain the
  Onsager relations.}
\end{remark}

Although the expressions \eref{Electrical Current 1}--\eref{Current Heat} for the electric and heat currents have a
direct physical interpretation in terms of $t_{ij}(E)$, we prefer to work in terms of the
quantities
\begin{equation}
\Gamma_{ij}(E) = M_i
\delta_{ij} - t_{ij}(E)~,
\end{equation}
where $\delta_{ij}$
denotes the Kronecker delta. For this, we use the relations \eref{Property
  Transmission} to rewrite the integrand in \eref{Electrical
  Current 1} as follows:
\begin{eqnarray}
\sum_{j} \left[f_i(E) t_{ji}(E) - f_j(E) t_{ij}(E)\right] &=& f_i(E) M_i - \sum_j f_j(E) t_{ij}(E) \nonumber\\
&=& \sum_j f_j(E) \underbrace{\left(M_i \delta_{ij} - t_{ij}(E)\right)}_{= \
  \Gamma_{ij}(E)}~.
\end{eqnarray}
Hence, one can rewrite the currents in the following form ($i\in \{\L, \R, 1,\dots,N\}$):
\begin{eqnarray}
I_i &=& \frac{e}{h} \sum_{j} \int_{0}^{\infty} f_j(E)
\Gamma_{ij}(E) \, dE~,\label{Electric Current Def}\\
J_i &=& \frac{1}{h} \sum_{j} \int_{0}^{\infty} f_j(E)
\Gamma_{ij}(E) (E - \mu_i) \, dE~.\label{Heat Current Def}
\end{eqnarray}

From \eref{Property Transmission} one immediately obtains
  the following properties:
\begin{equation}
\sum_{i} \Gamma_{ij}(E) = 0~, \hspace{2mm} \forall j \hspace{7mm} \mbox{and} \hspace{7mm} \sum_{j} \Gamma_{ij}(E) =
0~, \hspace{2mm} \forall i~.\label{Property Conductance}
\end{equation}

\begin{remark}
\textnormal{Note that $\Gamma_{ij}(E) \not = \Gamma_{ji}(E)$ in general.}
\end{remark}

Using the relations \eref{Property Conductance} one
  immediately sees that the expressions \eref{Electric Current
  Def}--\eref{Heat Current Def} for the currents satisfy the
  conservation of charge and energy, respectively, \ie
\begin{equation}\label{conservation currents}
\sum_{i} I_{i} = 0 \hspace{7mm} \mbox{and} \hspace{7mm} \sum_{i} \left(J_{i}
+ \frac{\mu_i}{e} I_{i}\right) = 0~.
\end{equation}

\begin{remark}
\textnormal{The second relation in \eref{conservation currents} emphasises
  the fact that, in general, energy is conserved but not heat, in the sense that $\sum_{i} J_{i} \not = 0$ in general. However, as we shall see in the next section (see \Remark~\ref{remark heat}), heat is
  conserved in the linear response approximation.}
\end{remark}

\subsection{The Strategy}\label{The strategy}

In \Section~\ref{The null currents condition}, we will fix the
values $(T_\L,\mu_\L)$ and $(T_\R,\mu_\R)$, and determine the parameters $(T_i,\mu_i)$, for $i = 1\dots,N$,
for which the \emph{self-consistency condition} is satisfied:
\begin{equation}
I_i = 0 \quad \mbox{and} \quad J_i = 0 \quad \mbox{for} \quad i =
1,\dots,N~.\label{Condition Zero Currents}
\end{equation}
Actually, this condition will be solved only in the linear response
approximation. Then we will work out the remaining electric and heat currents: $I_\L = - I_\R$ and
$J_\L = - J_\R$ (these equalities are satisfied in the linear response
regime; see \Remark~\ref{remark heat}). Observe that in this situation, charge and heat dissipation occurs mostly at the left and right boundaries of the system. Indeed, due to the self-consistency condition, the dissipation occurs at first order at the boundaries while at most as second order effects along the system.

Up to that point,
the system is described by any scattering matrix $S$ and has therefore
no specific geometry (see \Fig~\ref{The General System}). In order to
obtain a system with the \emph{linear}
geometry represented in \Fig~\ref{The Quantum System}, and to interpret $I_\L$ and $J_\L$ as the currents flowing through the system,
from left to right, we will consider that the \emph{global} scattering
matrix $S$ is given in terms of the \emph{local} (linearly ordered) scattering
matrices, $S^{(1)},\dots,S^{(N)}$, associated to the $N$ quantum
dots. 

In
\Section~\ref{Numerical analysis}, we will study numerically the properties of the
\emph{composite}
scattering matrix $S$, using random matrix theory (RMT), and show
that the currents through the chain of quantum dots with
self-consistent reservoirs are typically (in the sense of RMT) governed by
Ohm and Fourier laws.

\setcounter{equation}{0}
\section{General Properties}\label{General properties}

\begin{figure}[h!]
\begin{center}
\psfig{file=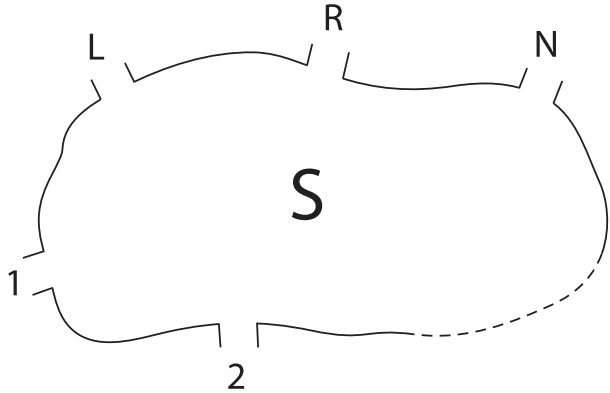,width=8cm}
\caption{The general system coupled to $N+2$ terminals.}
\label{The General System}
\end{center}\vspace{-2mm}
\end{figure}

In this section, we consider that the system is coupled to $N+2$ terminals
(we use $\L$ and $\R$ to be consistent with the next sections)
and suppose that its transport properties are given in terms of some
scattering matrix $S$ (see \Fig~\ref{The General System}). 

In the
linear response regime and under the assumption that the scattering
matrix $S$ does not depend on the energy, we will
determine the general transport properties of the system. All these properties are
expected to hold on physical grounds and actually some of them have recently
been obtained in full generality \cite{CA}. Nevertheless, in
this paper, we are only interested in some particular situations in which the
derivations of these properties are somehow simpler.

\subsection{Linear Transport Analysis}\label{Linear transport analysis}

In the remainder of this paper we shall always rely on the following.

\begin{assumption}
We assume that, for $i \in \{\L, \R, 1, \dots,
N\}$,
\begin{eqnarray}
T_i &=& T + \delta T_i~,\\
\mu_i &=& \mu + \delta \mu_i~,
\end{eqnarray}
where $T > 0$ and $\mu \in \mathbb{R}$ are some reference values, while $\delta T_i \in \mathbb{R}$
and $\delta \mu_i \in \mathbb{R}$ are some ``small'' perturbations.
For definiteness, we assume that all subsequent expressions are first
order expansions in $\delta T_i$ and $\delta \mu_i$.\label{Assumption Linear transport}
\end{assumption}

To study the transport, we consider the first order expansion of any distribution function:
\begin{equation}
g(E; T_j, \mu_j) = g(E; T, \mu) + \frac{\partial g}{\partial T_j}(E;
T, \mu) \ \delta T_j + \frac{\partial g}{\partial \mu_j}(E;
T, \mu)  \ \delta \mu_j~.
\end{equation}
Note that the three considered distribution functions
\eref{Maxwell-Boltzmann}--\eref{Bose-Einstein} are of the
form: 
\begin{equation}\label{form g}
g(E; T_j, \mu_j) = g\left(\frac{E-\mu_j}{k_\B
  T_j}\right)~.
\end{equation}
Hence, one can rewrite the derivatives of $g$, with respect to $T_j$ and $\mu_j$, in
terms of the derivative of $g$ with respect to $E$, and obtain the
following result:
\begin{equation}\label{First Order Exp}
f_j(E) \equiv f(E; T_j, \mu_j) = f(E; T, \mu) - \frac{\partial f}{\partial E}(E;
T, \mu) \ \left[\left(\frac{E-\mu}{T}\right) \delta T_j + \delta \mu_j
  \right]~.
\end{equation} 
Substituting this first order expansion into the expressions \eref{Electric Current
  Def}--\eref{Heat Current Def} for the currents, and using the relations
  \eref{Property Conductance}, one sees that all terms containing
  $f(E; T, \mu)$ vanish and one is left with 
\begin{eqnarray}
I_i &=& \sum_j L_{ij}^{(0)} \, \frac{\delta \mu_j}{e} + L_{ij}^{(1)}
\, \frac{\delta T_j}{T}~,\label{The electric current}\\
J_i &=& \sum_j L_{ij}^{(1)} \, \frac{\delta \mu_j}{e} + L_{ij}^{(2)}
\, \frac{\delta T_j}{T}~,\label{The heat
  current}
\end{eqnarray}
where 
\begin{eqnarray}
L_{ij}^{(0)} &=& -\frac{e^2}{h} \int_{0}^{\infty} \frac{\partial
  f}{\partial E}(E; T, \mu) \Gamma_{ij}(E) \, dE~,\label{transp 1}\\
L_{ij}^{(1)} &=& -\frac{e}{h} k_\B T \int_{0}^{\infty} \left(\frac{E-\mu}{k_\B T}\right) \frac{\partial
  f}{\partial E}(E; T, \mu) \Gamma_{ij}(E) \, dE~,\label{transp 2}\\
L_{ij}^{(2)} &=& -\frac{(k_\B T)^2}{h} \int_{0}^{\infty} \left(\frac{E-\mu}{k_\B T}\right)^2 \frac{\partial
  f}{\partial E}(E; T, \mu) \Gamma_{ij}(E)  \, dE~.\label{transp 3}
\end{eqnarray}

\begin{remark}\label{remark heat}
\textnormal{Using the relations \eref{Property
  Conductance}, one sees that the first order approximations \eref{The heat  current} for the heat currents
   satisfy
\begin{equation}
\sum_{i} J_{i} = 0~.\label{cons heat}
\end{equation}
In \Section~\ref{The null currents condition}, we will impose the
self-consistency condition,  $I_i = J_i = 0$ for $i =
1,\dots,N$, and the conservation laws \eref{conservation currents} and 
\eref{cons heat} will permit to interpret $I_\L = -I_\R$ and $J_\L = -J_\R$ as the
electric and heat currents flowing through the system (from left to
right).} 
\end{remark}

\subsection{The Transport Matrix}\label{The transport matrix}

Here we establish some basic properties of the transport matrix. From
the expressions \eref{The electric current}--\eref{The heat
  current} for the currents we are naturally led to the following definition.

\begin{definition}\label{def transport matrix}
We denote by $L$ the transport matrix:
\begin{equation*}
L = \matrice{L^{(0)} & L^{(1)}\\ L^{(1)} & L^{(2)}}~, \hspace{3mm} \mbox{where}  \hspace{3mm}  L^{(k)} =
  \matrice{L^{(k)}_{\L \L} & \L^{(k)}_{\L 1} & \dots & L^{(k)}_{\ LN} &
  \L^{(k)}_{\L \R}\\ L^{(k)}_{1 \L} & \L^{(k)}_{1 1} & \dots &
  L^{(k)}_{1 N} & L^{(k)}_{1 \R}\\ \vdots & \vdots &  \vdots & \vdots
  & \vdots \\ L^{(k)}_{N \L} &
  L^{(k)}_{N 1} & \dots & L^{(k)}_{N N} & L^{(k)}_{N \R} \\ L^{(k)}_{\R \L} &
  L^{(k)}_{\R 1} & \dots & L^{(k)}_{\R N} & L^{(k)}_{\R \R} }~.
\end{equation*}
\end{definition}

A quick glance at the expressions \eref{transp 1}--\eref{transp 3} for
the transport coefficients $L_{ij}^{(k)}$, remembering that $\Gamma_{ij}(E) =
M_i \delta_{ij} - t_{ij}(E)$, shows that the Onsager relations hold,
\ie if $S(E)$ is symmetric (for each $E > 0$), then
$L$ is symmetric. More generally, if the system satisfies the
micro-reversibility property $S_{ij,mn}(E,B) = S_{ji,nm}(E,-B)$, where $B$
is some applied magnetic field, then $L_{ij}(B) = L_{ji}(-B)$.

In general, the scattering matrix $S$ depends on the energy. Nevertheless, in this paper, we shall restrict our attention to the energy-independent situations. The assumption that $S$
does not depend on the energy is a good approximation in
some limiting cases (\eg looking at \eref{transp 1}--\eref{transp
  3} one sees that for electrons at low
temperature one may consider $S(E) \equiv S(E_\F)$, where  $E_\F$ is the Fermi energy, since $(-\partial f/\partial E)(E) \approx
\delta(E-E_\F)$) and leads to interesting consequences. In particular, we
will see that the relations among the transport coefficients, $L_{ij}^{(0)}$,
$L_{ij}^{(1)}$ and $L_{ij}^{(2)}$, will be very simple and will lead to some universal transport properties. For definiteness, we assume for the
rest of this paper that the following holds.

\begin{assumption}
The scattering matrix $S$ does not depend on the energy.\label{Assumption Gamma}
\end{assumption}

In order to simplify some derivations, without restricting our results
too much, it is convenient to assume the following.

\begin{assumption}
We assume that $t_{ij} \not = 0$ for all $i,j \in \{\L,\R, 1,\dots, N\}$.\label{Assumption Gamma2}
\end{assumption}

\begin{remark}
\textnormal{Recalling the properties \eref{Property Transmission}, one
  sees that the Assumption A\ref{Assumption Gamma2} implies that $t_{ij}
  \not = \min\{M_i,
  M_j\}$, for all $i,j \in \{\L,\R, 1,\dots, N\}$, and this permits to avoid total back-scattering (\ie $t_{ii} \not = M_i$ for all $i$). Although not optimal, the Assumption A\ref{Assumption Gamma2} also insures the existence and unicity of the solution of the self-consistently condition \eref{Condition Zero Currents}.}
\end{remark}

Under the assumption that the scattering matrix $S$ does not depend on the
energy, one sees in the expressions for the transport coefficients
\eref{transp 1}--\eref{transp 3} that some integrals of the following
form appear (with $n = 0, 1, 2$):
\begin{equation}
C(n) = - \int_{0}^{\infty} \left(\frac{E-\mu}{k_\B T}\right)^n \frac{\partial
  f}{\partial E}(E; T, \mu) \, dE~.
\end{equation}
More explicitly, 
\begin{equation}
C^{\rm MB}(n) = \int_{x_0}^{\infty}
  x^n e^{-x} \ dx \hspace{5mm} \mbox{and} \hspace{5mm}C^{\rm \pm}(n) =  \int_{x_0}^{\infty}
  \frac{x^n e^x}{(e^x \pm 1)^2} \ dx~,
\end{equation}
where the signs $+$ and $-$ correspond to the Fermi-Dirac and Bose-Einstein
cases, respectively, and $x_0 = -\mu/(k_\B T)$
is some reference parameter. Here, $x_0
\in \mathbb{R}$ in MB and FD, but $x_0
> 0$ in BE. 

\begin{remark}
\textnormal{Although $C(n)$ depends on $x_0$,
  we will see that the transport properties of the system are
  essentially independent of $x_0$. This explains why we do not
  write $C(n,x_0)$.}
\end{remark}

In terms of $C(n)$, one can write
\begin{eqnarray}
L_{ij}^{(0)} &=& \frac{e^2}{h} C(0)  \Gamma_{ij}~,\label{L0}\\
L_{ij}^{(1)} &=&  \frac{e}{h} k_\B T C(1) \Gamma_{ij}~, \label{L1}\\
L_{ij}^{(2)} &=& \frac{(k_\B T)^2}{h} C(2) \Gamma_{ij}~. \label{L2}
\end{eqnarray}
In Appendix~A, we show that $C(0)$, $C(1)$ and $C(2)$ are positive for
  all 
\begin{equation}
x_0 = -\frac{\mu}{k_\B T} \in \left\{\begin{array}{l}(-1,\infty)
\mbox{ in MB}\\ 
(-\infty,\infty) \mbox{ in FD}\\ (0,\infty) \mbox{ in BE}\end{array}\right.~.
\end{equation}
In the Maxwell-Boltzmann case, when $x_0 \in (-\infty,-1]$, we find
that $C^{\rm MB}(0)$ and $C^{\rm MB}(2)$ are positive, but $C^{\rm
  MB}(1)$ is non-positive. As one may easily check, all the subsequent
results also hold in this situation. Nevertheless, in order to be able to treat the three considered cases on the same footing we
make the following restricting assumption.
\begin{assumption}
In the Maxwell-Boltzmann case: $x_0 = -\mu/(k_\B T)\in (-1,\infty)$.\label{Assumption xo}
\end{assumption}

\begin{remark}
\textnormal{Note that the coefficient $C^{\rm BE}(0)$ diverges as
  $x_0 \rightarrow 0$. This may be interpreted as a
  Bose-Einstein condensation, and shows that our theory breaks down
  in describing properly the transport of Bose-Einstein condensates.}
\end{remark}

\begin{remark}\label{remarks FD}
\textnormal{In mesoscopic physics, one is usually interested in electronic
  transport at low
  temperature, \ie $x_0
  = -\mu/(k_\B T) \rightarrow -\infty$. Observe that in this limit $C^{\rm FD}(1) \rightarrow 0$ and consequently $L_{ij}^{{\rm
FD}(1)} \rightarrow 0$. In order to have non-zero transport
  coefficients $L_{ij}^{{\rm
  FD}(1)}$ in this regime, one usually considers the first order term in the Taylor expansion, $\Gamma_{ij}(E) = \Gamma_{ij}(\mu) +
  \Gamma'_{ij}(\mu) (E-\mu) + O((E-\mu)^2)$, which leads to $L_{ij}^{{\rm
FD}(1)} \sim \Gamma'_{ij}(\mu)$.}
\end{remark}

\begin{remark}
\textnormal{Observe that the transport coefficients
$L_{ij}^{(k)}$ depend on the properties of the quantum dots through
the scattering matrix $S$. Here we can
see why the assumption A\ref{Assumption Gamma2} is useful. Indeed, suppose that
$t_{ii} = M_i$ for some $i$ (total reflection). Then
$t_{ij} = 0$ for all $j \not = i$ and consequently $L^{(k)}_{ij} = 0$ for all $j$. In other words, in such a situation the $i$-th reservoir is
  disconnected from the rest of the system.}
\end{remark}

Looking at the relations \eref{L0}--\eref{L2}, one sees that the transport
coefficients are related as follows:
\begin{equation}
 L_{ij}^{(0)} = \frac{e^2}{h} C(0) \Gamma_{ij}~,
\hspace{5mm}  L_{ij}^{(1)} = Q_1 L_{ij}^{(0)}  \hspace{5mm}
\mbox{and} \hspace{5mm}
L_{ij}^{(2)} = Q_2 L_{ij}^{(0)}~,
\end{equation}
where
\begin{equation}
Q_1 =\frac{k_\B T}{e} \frac{C(1)}{C(0)} > 0 \hspace{5mm}
\mbox{and} \hspace{5mm} Q_2 = \frac{k_\B^2 T^2}{e^2} \frac{C(2)}{C(0)} > 0~.\label{L4L1}
\end{equation}

\begin{remark}\label{Autre remarks FD2}
\textnormal{The quantity $Q_1$ is related to the Seebeck coefficient
  (see \Remark~\ref{Rem Seebeck}). In the Fermi-Dirac
  case, $\lim_{x_0 \rightarrow  -\infty} Q_2/T^2 = (\pi^2/3) (k_\B/e)^2$ is called the Lorentz
number.}
\end{remark}

Recalling that $\Gamma_{ij} =
M_i \delta_{ij} - t_{ij}$, one immediately obtains the following properties ($k = 0,1,2$):
\begin{equation} \label{Property Lij}
L^{(k)}_{ij} \ \left\{\begin{array}{ll} > 0 &
\mbox{if } \ i
= j\\
< 0 &
\mbox{if } \ i
\not = j
\end{array}\right.~,
\end{equation} 
and
\begin{equation}\label{Property Lij2}
\sum_{i} L^{(k)}_{ij} = 0~, \hspace{2mm} \forall j \hspace{7mm} \mbox{and} \hspace{7mm} \sum_{j} L^{(k)}_{ij} =
0~, \hspace{2mm} \forall i~.
\end{equation}

As we shall see later on, the following ratio, $\mathcal{R}$, will play an important
role. Using standard techniques, we show in Appendix~B that (for all $x_0$):

\begin{equation}\label{Ratio R}
\mathcal{R} \equiv \frac{Q_1}{\sqrt{Q_2}} =
\frac{C(1)}{\sqrt{C(0) \cdot C(2)}} \in (0,1)~.
\end{equation}

\begin{remark}
\textnormal{In terms of the matrix
\begin{equation}
C = \matrice{C(0) & C(1)\\C(1) & C(2)},
\end{equation}
the inequality $0 < \mathcal{R} < 1$ is equivalent to $\det(C) = C(0)
C(2) - C(1)^2 > 0$. }
\end{remark}

\subsection{The Entropy Production}\label{The entropy production}

In this subsection, we show that the transport matrix $L$ is real
positive semi-definite. In other words, we show that the entropy
production rate is non-negative:
\begin{equation}\label{entr prod}
\sigma_s = \sum_{i,j=1}^{2N+4}  L_{ij} V_i V_j \geq 0~,
\end{equation}
where the thermodynamic forces are 
\begin{equation}
V_i = \left\{ \begin{array}{ll}\delta \mu_{i-1}/e & \mbox{ if } \; \, i = 1,
  \dots, N+2\\ \delta T_{i-(N+3)}/T &  \mbox{ if } \; \; i = N+3, \dots, 2N+4\end{array}\right. ~.
\end{equation}
Here, we have set $\L = 0$ and $\R = N+1$. Let $X_i = \delta \mu_i/e$ and $Z_i = \sqrt{Q_2} \ \delta
T_i/T$, for $i \in \{0, 1, \dots, N, N+1\}$. Then, in Appendix~C, we
show that one can rewrite the entropy production rate as
follows:
\begin{equation}\label{sigmas expression}
\sigma_s = \sum_{\scriptsize\begin{array}{c}
    i,j = 0\\ i < j \end{array}}^{N+1} (-L^{(0)}_{ij}) \ I_{ij}~,
\end{equation}
where
$$
I_{ij} = (X_i - X_j)^2
  + (Z_i - Z_j)^2 -  2 \mathcal{R} C_{ij}~, \hspace{5mm} C_{ij} = X_i Z_j + X_j Z_i - X_i Z_i - X_j Z_j~.
$$
Observe now that $(-L^{(0)}_{ij}) > 0$ for all $i \not = j$. Therefore, to
show that $\sigma_s \geq 0$ it is
sufficient to show that $I_{ij} \geq 0$ for all $i \not = j$. As we shall
see, it will be crucial that the ratio  $\mathcal{R}$
satisfies $0 < \mathcal{R} < 1$. Assume first that $C_{ij} \leq 0$, then ($
\mathcal{R} > 0$)
\begin{equation}\label{In1a}
I_{ij} \geq (X_i - X_j)^2 + (Z_i - Z_j)^2 \geq 0~.
\end{equation}
Assume next that $C_{ij} > 0$, then  ($\mathcal{R} < 1$)
\begin{equation}
I_{ij} >  (X_i - X_j)^2 + (Z_i - Z_j)^2  - 2 C_{ij} = (X_i - X_j + Z_i - Z_j)^2 \geq 0~.
\end{equation}
This ends the proof that $\sigma_s \geq 0$.

\begin{remark}
\textnormal{The term $(X_i - X_j)^2$ accounts for the entropy
  production due to the electric current, and it is well known in
  mesoscopic physics (see \eg \cite{Buttiker-Symmetry}). The term
  $(Z_i - Z_j)^2$ is related to the heat current, and one sees that
  there is also a thermoelectric term, $-2 \mathcal{R} C_{ij}$, which
  might take positive and negative values.}
\end{remark}

\subsection{Equilibrium and Non-Equilibrium States}\label{Equilibrium
  and non-equilibrium states}

Here we present some equivalent characterizations of the equilibrium
and non-equilibrium states of the multi-terminal system. We start with the following.

\begin{definition}
We say that the system is at \emph{equilibrium} if 
\begin{equation*}
T_\L = T_1 = \dots = T_N = T_\R  \hspace{5mm}\mbox{ and } \hspace{5mm}
\mu_\L = \mu_1 = \dots = \mu_N = \mu_\R~.
\end{equation*}
Otherwise, we say that the system is \emph{out of equilibrium}. 
\end{definition}

Now, as one expects, the system is at equilibrium only in the
situations in which all the electric and heat currents vanish, \ie
\begin{equation*}
\{\mbox{System is at equilibrium}\} \iff \{I_i = 0 \hspace{3mm}
\mbox{and} \hspace{3mm} J_i = 0, \hspace{3mm} \forall i\}~.
\end{equation*}
This equivalence is obtained by writing the no-current condition, $I_i
= J_i = 0$ for all $i \in \{\L, \R, 1,\dots,N\}$, in a matrix form and
by using the
properties of the transport coefficients $L^{(k)}_{ij}$ as well as the
fact that $0 < \mathcal{R} < 1$. The details are presented in Appendix~D.

Also, one can characterize the equilibrium
and non-equilibrium  states of the system in terms of the entropy
production:
$$
\left\{\hspace{-1mm}\begin{array}{cc}\mbox{System is at}\\\mbox{equilibrium}\end{array}\hspace{-1mm}\right\} \iff \sigma_s = 0 \hspace{5mm} \mbox{and}
\hspace{5mm} \left\{\hspace{-1mm}\begin{array}{cc}\mbox{System is}\\\mbox{out of
  equilibrium} \end{array}\hspace{-1mm}\right\}\iff \sigma_s > 0~.
$$
These equivalences easily follow from the expression \eref{sigmas
  expression} for $\sigma_s$ and the inequality \eref{In1a}. The details can be found in Appendix~D.

\section{The Self-Consistency Condition}\label{The null currents condition}

We now turn to the resolution of the self-consistency condition
\eref{Condition Zero Currents}. Using the expressions \eref{The
  electric current}--\eref{The heat current} one can
rewrite the self-consistency condition as
follows ($i = 1,\dots,N$):
\begin{eqnarray}
\sum_{j = 1}^{N} \left(L_{ij}^{(0)} \, \frac{\delta \mu_j}{e} + L_{ij}^{(1)}
\, \frac{\delta T_j}{T}\right) &=& - \sum_{j = \L,\R} \left(L_{ij}^{(0)} \, \frac{\delta \mu_j}{e} + L_{ij}^{(1)}
\, \frac{\delta T_j}{T}\right)~,\label{Condition1}\\
\sum_{j = 1}^{N} \left(L_{ij}^{(1)} \, \frac{\delta \mu_j}{e} + 
L_{ij}^{(2)}
\, \frac{\delta T_j}{T}\right) &=& - \sum_{j = \L,\R} \left(L_{ij}^{(1)} \, \frac{\delta \mu_j}{e} + L_{ij}^{(2)}
\, \frac{\delta T_j}{T}\right)~.\label{Condition2}
\end{eqnarray}
We recall that we are given the values $(T_\L, \mu_\L)$ and
$(T_\R, \mu_\R)$, so that the right-hand side of the above equations are
supposed to be known ($T_j = T + \delta T_j$ and
$\mu_j = \mu + \delta \mu_j$). Therefore, the self-consistency
condition constitutes a set of $2N$ equations for
the $2N$ unknown variables $T_1, \dots, T_N$ and $\mu_1, \dots,
\mu_N$.

To solve these equations, it is convenient to introduce the
vectors $X, Y \in \mathbb{R}^{N}$ and the $N\times N$ matrix $L^{(0)}_{\rm C}$
defined by ($i,j = 1,\dots,N$):
\begin{equation}
X_j = \frac{\delta \mu_j}{e}~, \hspace{5mm} Y_j = \frac{\delta
  T_j}{T}~,  \hspace{5mm}
  \mbox{and} \hspace{5mm} (L^{(0)}_{\rm C})_{ij} = L^{(0)}_{ij}~.
\end{equation} 

\begin{remark}
\textnormal{Recalling that $L^{(0)}$ is the $(N+2) \times (N+2)$ matrix defined in
\Definition~\ref{def transport matrix}, one sees that $L^{(0)}_{\rm
  C}$ is the reduced matrix associated to the ``central''
terminals $1,\dots,N$.} 
\end{remark}

We also define for $\ell = \L, \R$:
\begin{equation}
X_\ell = \frac{\delta \mu_\ell}{e}~, \hspace{5mm} Y_\ell = \frac{\delta
  T_\ell}{T}~, \hspace{5mm}
  \mbox{and} \hspace{5mm} D_\ell = \matrice{L^{(0)}_{1 \ell} \\ \vdots
  \\ L^{(0)}_{N \ell}}~.
\end{equation} 
Then the equations \eref{Condition1}--\eref{Condition2} can be rewritten as
follows:
\begin{eqnarray}
L^{(0)}_{\rm C} \left(X + Q_1 Y\right) &=& - \sum_{\ell = \L, \R}
\left(X_\ell +  Q_1 Y_\ell \right) \ D_\ell~, \label{Eq XY1}\\
L^{(0)}_{\rm C} \left(Q_1 X + Q_2 Y\right) &=& - \sum_{\ell = \L, \R}
\left(Q_1 X_\ell +  Q_2 Y_\ell \right) \ D_\ell~.\label{Eq XY2}
\end{eqnarray} 
Since $\mathcal{R} \not = 1$, one has $Q_2 \not =
(Q_1)^2$. As a consequence, if one takes $(Q_2/Q_1)$ times
\eref{Eq XY1} and subtract $\eref{Eq XY2}$ and similarly if one takes $(-1/Q_1)$ times \eref{Eq XY2} and add \eref{Eq
  XY1}, one obtains
\begin{equation}
L^{(0)}_{\rm C} X = - \sum_{\ell = \L, \R} D_\ell X_\ell \hspace{5mm}
  \mbox{and} \hspace{5mm} L^{(0)}_{\rm C} Y = - \sum_{\ell = \L, \R} D_\ell
  Y_\ell~. \label{Equ MXY}
\end{equation}
We see here that the equations
\eref{Eq XY1}--\eref{Eq XY2} decouple and that the equations for
the chemical potentials and for the temperatures are actually
identical. In order to proceed, we need to know that the matrix
$L^{(0)}_{\rm C}$ is invertible. This fact is proved in Appendix~E by showing
that $L^{(0)}_{\rm C}$ is real positive definite. In particular, this shows
that the self-consistency condition \eref{Condition Zero Currents} has a unique solution. 

By inverting the matrix $L^{(0)}_{\rm C}$, we obtain the following
self-consistent temperatures and chemical
potentials ($i = 1,\dots,N$):
\begin{eqnarray}
T_i &=& T_\L +  A_i \,
(T_\R - T_\L)~,\label{Eq2 Profile}\\
\mu_i &=& \mu_\L +  A_i \,
(\mu_\R - \mu_\L)~,\label{Eq1 Profile}
\end{eqnarray}
where
\begin{equation}
A_i = \sum_{j = 1}^{N} (\Gamma_{\rm C}^{-1})_{ij} \ t_{j \R}~, \hspace{5mm}
\mbox{with}  \hspace{5mm} \Gamma_{\rm C} = (\Gamma_{ij})_{i,j = 1}^{N}~. \label{Eq2 Profileb}
\end{equation}
The details about their derivation are given in Appendix~F. 

By substituting these expressions into the relations \eref{The
  electric current}--\eref{The heat current} for the currents, with $i = \L$, and by
  using the properties of the transport coefficients $L^{(k)}_{ij}$,
  one deduces the (average) currents flowing through the system (from left to right):
\begin{eqnarray}
I_\L &=& \sigma_0 \left( \frac{\mu_\R - \mu_\L}{e}\right) + \sigma_1
\left( \frac{T_\R - T_\L}{T}\right)~,\label{Current I}\\
J_\L &=& \sigma_1 \left(\frac{\mu_\R - \mu_\L}{e}\right) + \sigma_2  \left(\frac{T_\R - T_\L}{T}\right)~,\label{Current J}
\end{eqnarray}
where
\begin{equation}
\sigma_0 = L^{(0)}_{\L\R} + \sum_{j = 1}^{N} A_j \ L^{(0)}_{\L j}~,
\hspace{5mm} \sigma_1 = Q_1 \sigma_0 \hspace{5mm} \mbox{and} \hspace{5mm} \sigma_2 = Q_2 \sigma_0~. \label{Sigma1}
\end{equation}

\begin{remark}
\textnormal{The corresponding expressions to \eref{Eq2
    Profile}--\eref{Sigma1} in the case of
  electronic transport at low temperature ($T_i = 0$ and $J_i = 0$ for
  $i \in \{\L, \R, 1, \dots N\}$) were obtained in \cite{Amato}.}
\end{remark}

\begin{remark}\label{Rem Seebeck}
\textnormal{Setting $I_\L = 0$, one deduces the thermoelectric field
  $\mathcal{E} = - \nabla V =
  \mathcal{S} \nabla T$, where $\mathcal{S} = Q_1/T = k_\B/e \cdot
  C(1)/C(0) > 0$ is the thermopower or Seebeck coefficient. Note that
  $\mathcal{S}$ does not depend on $N$.  For further discussions see \eg \cite{Streda,BeenakkerThermoPower,StaringOscillations,Molenkamp,Godijn,Mathias,Kato}. }
\end{remark}

Using the relations \eref{Eq2 Profile}--\eref{Eq1 Profile}, one may
rewrite the currents \eref{Current I}--\eref{Current J} as follows:
\begin{eqnarray}
I_\L &=& \sum_{j} L^{(0)}_{\L j}  \left( \frac{\mu_j -
  \mu_\L}{e}\right) +  L^{(1)}_{\L j} \left(\frac{T_j - T_\L}{T}\right)~,\\
J_\L &=& \sum_{j} L^{(1)}_{\L j}  \left( \frac{\mu_j -
  \mu_\L}{e}\right) +  L^{(2)}_{\L j} \left(\frac{T_j - T_\L}{T}\right)~.
\end{eqnarray}

Although the multi-terminal system considered so far has no specific
  geometry (see \Fig~\ref{The
  General System}), we shall nevertheless use the word ``profile'' for the arrangements $\mu_1, \dots,
  \mu_N$ and $T_1, \dots, T_N$. This terminology will be fully justified in
  \Section~\ref{Numerical analysis}, where we will restrict our attention to the linear
  chain of quantum dots (\Fig~\ref{The Quantum System}).

\begin{remark}
\textnormal{Note that the temperature and chemical potential profiles
  are decoupled, \ie $T_i$ only depends on $(T_\L, T_\R)$ and
  $\mu_i$ only depends on $(\mu_\L, \mu_\R)$. On the other
  hand, since $Q_1 > 0$, there can be a heat current due to a gradient of
chemical potential ($T_\L = T_\R$), and reciprocally, there can
  be an electric current due to a gradient of temperature ($\mu_\L =
  \mu_\R$).}
\end{remark}

\begin{remark}\label{remark on B}
\textnormal{We see that the form of the profiles is given by the coefficients $A_1, \dots, A_N$. Hence, in what follows we will
  use the word ``profile'' to refer to these coefficients. Looking at
  the expression \eref{Eq2 Profileb} for $A_i$ one sees that it does not
  depend on the distribution function \eref{Maxwell-Boltzmann}--\eref{Bose-Einstein} describing the reservoirs. In this sense, we
  say that the profile is \emph{universal}. One may wonder whether this is true for any
  distribution function $f$. It seems not to be the case. Indeed, the
  key ingredients to obtain the preceding results are: (i) the specific form \eref{form g} of $f$ and (ii) the ratio
$\mathcal{R}$, defined in \eref{Ratio R}, must satisfy $0 < \mathcal{R} <
1$. Basically, we expect that any
distribution function $f$ satisfying (i)-(ii) will lead to the same
  results. Note that universal temperature profiles were also found for a quantum harmonic chain
  coupled at both ends to two phonon reservoirs at different
  temperatures \cite{Saito}.}
\end{remark}

\begin{remark}\label{remark on B}
\textnormal{Although the profiles are the same in the three considered
  situations, one can distinguish between the classical and quantum chain of quantum dots. Indeed, assume we are
  given the scattering matrices, $S^{(1)}, \dots, S^{(N)}$, associated
  to the $N$ quantum dots. Then, as explained in detail in Appendix~H,
  one can compose them into a global scattering matrix $S$, out of
  which one extracts the transmission probabilities $t_{ij}$ (see \eref{transmission prob}). This is the natural \emph{quantum} way of
  working.} 

\textnormal{In order to extract the interference effects, one may first
  compute the probability
matrices, $P^{(1)}, \dots, P^{(N)}$, defined by $P^{(k)}_{ij;mn} = |S^{(k)}_{ij;mn}|^2$, then compose them into a global probability matrix
  $P$ and finally set $t_{ij} = \sum_{m=1}^{M_i}
\sum_{n=1}^{M_j} P_{ij;mn}$. This way of computing $t_{ij}$ will be referred to as \emph{classical}.} 

\textnormal{As one may check, the composition law of the probability
matrices is the same as for the scattering matrices, \ie all
  expressions in Appendix~H hold with $P$ instead of $S$. In this
  sense, we will study in \Section~\ref{Numerical analysis} the classical versus
quantum situations (see also \cite{ShapiroB,CahayArray}).}
\end{remark}

In Appendix~G, we derive an interesting alternative expression
for the coefficients $A_i$. We obtain ($i = 1, \dots, N$)
\begin{equation}\label{Prop B}
A_i = \frac{\sum_{j = 1}^{N} (-1)^{i+j} \det\left(\Gamma_{\rm C}(j,i)\right)
  \ t_{j \R}}{\sum_{j = 1}^{N} (-1)^{i+j} \det\left(\Gamma_{\rm C}(j,i)\right)
  \ [t_{j \L} + t_{j \R}]}~,
\end{equation}
where $\Gamma_{\rm C}(j,i)$ denotes the $(j,i)$ minor of $\Gamma_{\rm  C}$.

\begin{remark}\label{Remark on B1}
\textnormal{As an illustration, let us apply the relation \eref{Prop
    B} in the case $N = 1$. One has
\begin{equation}\label{Eq on B1}
A_1 = \frac{t_{1 \R}}{t_{1 \L} + t_{1 \R}} \hspace{5mm}
\Longrightarrow \hspace{5mm}\mu_1 = \frac{t_{1\L} \ \mu_\L + t_{1\R} \
  \mu_\R}{t_{1 \L} + t_{1 \R}}~.
\end{equation}
This relation for
$\mu_1$ corresponds to the one obtained in \cite{Buttiker-Symmetry} for a
three-terminal conductor under the conditions $T_\L = T_1 = T_\R =
0$ (which implies $J_1 = 0$, since $\delta T_j = 0$ and $L_{ij}^{{\rm FD} (1)} = 0$ in this
case) and $I_1 = 0$.}
\end{remark}

An important consequence of \eref{Prop
    B} is ($i = 1, \dots, N$):
\begin{equation}\label{Proposition B Sigma}
A_i \in (0,1)~.
\end{equation}
This implies that the temperatures
  and chemical potentials of the reservoirs,  connected self-consistently to the system, are bounded between the boundary
  values $T_\L, T_\R$ and $\mu_\L, \mu_\R$, respectively. In other
  words, for $i =
  1, \dots,N$, one has
$$
\min\{ T_\L, T_\R \} \leq T_i \leq \max\{ T_\L, T_\R \}
\hspace{5mm}\mbox{and} \hspace{5mm} \min\{ \mu_\L, \mu_\R \} \leq \mu_i \leq \max\{ \mu_\L, \mu_\R \}~.
$$
\begin{remark}
\textnormal{The property \eref{Proposition B Sigma} also implies that the
  self-consistent parameters $T_1, \dots, T_N$ and $\mu_1, \dots, \mu_N$
  are increasing functions of $T_\L$, $T_\R$, and $\mu_\L$, $\mu_\R$, respectively.}
\end{remark}
Moreover, looking at the expressions \eref{Sigma1} for the coefficients $\sigma_0$,
$\sigma_1$ and $\sigma_2$, and
recalling that $L^{(0)}_{ij} < 0$ if $i \not = j$, $Q_1 > 0$ and $Q_2 > 0$, one deduces
that for $k = 0,1,2$:
\begin{equation}\label{Proposition B Sigma2}
\sigma_k < 0~.
\end{equation}

\begin{remark}
\textnormal{There are some simple consequences of \eref{Proposition B
  Sigma2} with respect to \eref{Current I}--\eref{Current J}.  (i) If
  $\mu_\L = \mu_\R$, then the heat current flows from the hot reservoir
  to the cold one, and similarly (ii) if $T_\L = T_\R$, then the
  electric current goes from the high chemical potential to
  the lower one. Note, however, that by setting $\mu_\L, \mu_\R$ and
  $T_\L, T_\R$ to appropriate values there might be a heat current going from the cold
  reservoir to the hot one. (A similar statement holds for the electric current.)}
\end{remark}

\subsection{Ohm and Fourier Laws}\label{Ohm and Fourier laws theory}

From the relations \eref{Current I}--\eref{Current J} for the
electric and heat currents, one can find explicit expressions for the
electric and heat conductances. If $T_\L = T_\R$, then setting $\mu = e V$, where $V$ denotes the
electric potential, one obtains 
\begin{equation}\label{Eq Ohm global1}
I_\L = - G_{\rm e} \ (V_\R - V_\L)~,
\end{equation}
where the electric conductance $G_{\rm e}$ is given by
\begin{equation}
G_{\rm e} = - \sigma_0~.
\end{equation}
If $I_\L = 0$, one deduces 
\begin{equation}\label{Eq Fourier global1}
J_\L = - G_{\rm h} \ (T_\R - T_\L)~,
\end{equation}
where the heat conductance $G_{\rm h}$ is given by
\begin{equation}
G_{\rm h} = \frac{\sigma_1^2 - \sigma_0 \sigma_2}{\sigma_0 T} = -
\frac{(Q_2 - Q_1^2)}{T} \ \sigma_0 = - \frac{k_\B^2 T}{e^2} \ \frac{C(0)
  C(2) - C(1)^2}{C(0)^2} \ \sigma_0~.
\end{equation}
Recalling that $0 < \mathcal{R} < 1$, one sees that $Q_2 >
(Q_1)^2$. Consequently, since $\sigma_0 < 0$, one deduces that
the  electric and heat conductances are positive: 
\begin{equation}
G_{\rm e} > 0 \hspace{5mm}  \mbox{and} \hspace{5mm} 
G_{\rm h} > 0~.
\end{equation}

Assume now that the system has the
linear geometry represented in \Fig~\ref{The Quantum System} and let
$a$ denote the length of the quantum dots and $\Sigma$ denote the width of the openings
connecting any two dots (\ie the effective cross
section of the system). Since $L=Na$ corresponds to the length of the system
made of $N$ quantum dots, one may introduce the global gradients:
\begin{equation}\label{global gradients}
\nabla V = \frac{V_\R
- V_\L}{L} \hspace{7mm} \mbox{and}  \hspace{7mm} \nabla T = \frac{T_\R
- T_\L}{L}~.
\end{equation}
Then, under the same conditions as in \eref{Eq Ohm global1} and \eref{Eq
Fourier global1}, the current \emph{densities}, $\mathcal{I}_\L = I_\L /\Sigma$ and
$\mathcal{J}_\L = J_\L /\Sigma$, are given by the \emph{global} Ohm and Fourier laws:
\begin{equation}\label{Eqs global}
\mathcal{I}_\L = - \kappa_{\rm e} \nabla V \hspace{5mm} \mbox{and} \hspace{5mm} \mathcal{J}_\L = - \kappa_{\rm h} \nabla T~,
\end{equation}
where the electric and heat conductivities, $\kappa_{\rm e}$ and
$\kappa_{\rm h}$,
are given by
\begin{equation}\label{Eqs cond global}
\kappa_{\rm e} = \frac{L G_{\rm e}}{\Sigma} > 0 \hspace{5mm} \mbox{and} \hspace{5mm}
\kappa_{\rm h} = \frac{L G_{\rm h}}{\Sigma} > 0~.
\end{equation}

\begin{remark}
\textnormal{In the Fermi-Dirac situation, one has
$$
\lim_{x_0 \rightarrow 0} Q_1 = 0 \hspace{2mm} \mbox{and}
  \hspace{2mm} \lim_{x_0
  \rightarrow 0} Q_2 = \frac{\pi^2}{3} \left(\frac{k_\B}{e}\right)^2 T^2
  \hspace{2mm} \Longrightarrow  \hspace{2mm} \lim_{x_0 \rightarrow 0} \frac{\kappa_{\rm h}}{\kappa_{\rm e} T} = \frac{\pi^2}{3} \left(\frac{k_\B}{e}\right)^2~.
$$
The last relation is the Wiedemann-Franz
law giving the Lorentz number.}
\end{remark}

In a similar manner, one may introduce the following local gradients:

\begin{equation}
\nabla V(i) = \frac{V_{i+1} - V_i}{a} \hspace{5mm} \mbox{and} \hspace{5mm}
\nabla T(i) = \frac{T_{i+1} - T_i}{a}~.
\end{equation}
Let $\mathcal{I}(i) = \mathcal{I}_\L$ and $\mathcal{J}(i) =
 \mathcal{J}_\L$, where $i \in \{1, \dots, N-1\}$, denote the net local current densities flowing from the
 $i$-th quantum dot into the $(i+1)$-th quantum dot. Then, using the equations \eref{Eq2
  Profile}--\eref{Eq1 Profile} and  \eref{Eqs global}, one obtains the \emph{local} Ohm and Fourier laws:
\begin{equation}\label{Local OF}
\mathcal{I}(i) = - \kappa_{\rm e}(i) \nabla V(i) \hspace{5mm} \mbox{and}
\hspace{5mm} \mathcal{J}(i) = - \kappa_{\rm h}(i) \nabla T(i)~,
\end{equation}
where the \emph{local} electric and heat conductivities are given by
\begin{equation}
\kappa_{\rm e}(i) = \frac{\kappa_{\rm e}}{N(A_{i+1} - A_i)} \hspace{5mm}
\mbox{and} \hspace{5mm} \kappa_{\rm h}(i) = \frac{\kappa_{\rm h}}{N(A_{i+1} - A_i)}~.
\end{equation}

\begin{remark}
\textnormal{Although the global conductivities, $\kappa_{\rm e}$ and $\kappa_{\rm h}$, are always positive, we will see, in
\Section~\ref{Numerical analysis}, that the local conductivities,
$\kappa_{\rm e}(i)$ and $\kappa_{\rm h}(i)$, may take negative values.}
\end{remark}

The coefficients $\sigma_0$,
$\sigma_1$ and $\sigma_2$, the conductances $G_{\rm e}$ and
$G_{\rm h}$, and the conductivities $\kappa_{\rm e}$ and $\kappa_{\rm h}$ depend on the distribution function $f$ describing the
reservoirs. Nevertheless, they are all proportional to $\sigma_0$, 
with some multiplicative factor independent of $N$. The important
observation is that the dependence on the nature of the particles is
entirely contained in these multiplicative factors. This permits to introduce the
following universal quantity.
\begin{definition}\label{universal conductance}
We define the \emph{universal conductivity} as
\begin{equation}\label{universal conductance1}
\kappa(N) = -\frac{h}{e^2 C(0)} N \sigma_0  = N \left[t_{\L\R} +
  \sum_{i = 1}^{N} \sum_{j = 1}^{N} t_{\L j} \ (\Gamma_{\rm
  C}^{-1})_{ji} \ t_{i \R}\right]~.
\end{equation}
\end{definition}

\begin{remark}\label{Remark on Sigma1}
\textnormal{As an illustration, let us consider the case $N=1$. One has
\begin{equation}\label{equ kappa1}
\kappa(1) = t_{\L \R} + \frac{t_{\L 1} \ t_{1 \R}}{t_{1 \L} + t_{1 \R}}~.
\end{equation}
As expected, the universal conductivity is the sum of two contributions: a
  direct coherent transmission  and an indirect incoherent transmission through the intermediate
  terminal. This result is well known in mesoscopic physics
  \cite{Buttiker-Symmetry}, and one sees that the relation
  \eref{universal conductance1} is the generalisation to $N$
  self-consistent reservoirs.}
\end{remark}

Looking at the expression \eref{universal conductance1} for the
universal conductivity, one sees that $\kappa(N) > 0$, since $\sigma_0 <
0$ and $C(0) > 0$. Using moreover the
relations \eref{Property Transmission} and $t_{\L \R} < \min\{M_\L,
M_\R\}$, one finds the following \emph{optimal} bounds:
\begin{equation}
0 < \kappa(N) < N \cdot \min\{M_\L, M_\R\}~.\label{bounds}
\end{equation}

\setcounter{equation}{0}
\section{Numerical Simulations}\label{Numerical analysis}
\subsection{The Chain of Quantum Dots}

To prove the validity of the \emph{global} Ohm and Fourier laws in our model, it remains
to show that the transport is \emph{normal}, that is, when the system becomes macroscopic the global electric and heat
conductivities, $\kappa_{\rm e}$ and  $\kappa_{\rm h}$, become intrinsic properties of the macroscopic system and consequently become independent of its length $L = N a$. More precisely, one has to prove that the global electric and heat
conductivities converge to finite limits as $L \rightarrow \infty$. Looking at
the relations \eref{Eqs cond global} and \eref{universal
  conductance1}, one sees that $\kappa_{\rm e}$ and $\kappa_{\rm h}$ are both proportional to 
$\kappa(N)$, with some multiplicative factor independent of $N$, and consequently it is sufficient to
investigate the limit of $\kappa(N)$ as $N \rightarrow \infty$. 

As explained previously, the multi-terminal system considered
so far has no specific geometry (\Fig~\ref{The General System}), and therefore $\kappa(N)$ will not converge as $N \rightarrow~\infty$ in general. In order  obtain a finite
limit for $\kappa(N)$ as $N \rightarrow
\infty$ we shall proceed as follows. 

Let $\left(S^{(k)}\right)_{k = 1}^{\infty}$ be an infinite sequence of
scattering matrices, each one being associated to a quantum dot, and let $S_N$ denote the \emph{composite} global
scattering matrix associated to $\left(S^{(1)},\dots,S^{(N)}\right)$. The
details concerning the construction of $S_N$ are given in
Appendix~H. Having set-up the linear geometry represented in
\Fig~\ref{The Quantum System}, one may expect that
$\kappa(N)$ will converge to a finite value, $\kappa^{\infty}$, as $N
\rightarrow \infty$, and this is what we found in \emph{typical} numerical
experiments (see below). Unfortunately, 
the composite scattering matrix, $S_N$, is given in a rather
cumbersome form in terms of $S^{(1)},\dots,S^{(N)}$, which prevents
us from proving the existence of $\lim_{N \rightarrow \infty}
\kappa(N)$. 

Nevertheless, we will see by using numerical simulations
under random matrix theory, that the universal conductivity $\kappa(N)$
typically admits a finite limit, $\langle  \kappa^{\infty}  \rangle$, as $N \rightarrow
\infty$. This shows that the \emph{global} Ohm and Fourier laws hold on statistical
average, and we may write 
\begin{equation}\label{eq Ge}
\langle \kappa^{\infty}_{\rm e}  \rangle = \frac{a}{\Sigma} \ \frac{e^2 C(0)}{h} \ \langle \kappa^{\infty}  \rangle
\end{equation}
and
\begin{equation}\label{eq Gh}
\langle  \kappa^{\infty}_{\rm h}  \rangle =  \frac{a}{\Sigma} \ \frac{k_\B^2
  T}{h} \ \frac{C(0) C(2) - C(1)^2}{C(0)} \
\langle  \kappa^{\infty}  \rangle~.
\end{equation}

Moreover, we will see that typically the profile,
$A_1,\dots,A_N$, becomes linear as the number of dots $N \rightarrow
\infty$. Introducing the variable $x = i/(N+1) \in [0,1]$, this means that in
the thermodynamic limit $N \rightarrow
\infty$, one has 
\begin{equation}
\langle 
A(x)  \rangle = x~,
\end{equation}
and consequently 
\begin{equation}
\langle V(x) \rangle = V_\L + (V_\R - V_\L) \ x \hspace{5mm} \mbox{and}  \hspace{5mm}
\langle T(x) \rangle = T_\L + (T_\R - T_\L) \ x~.
\end{equation}
Hence, the \emph{local} Ohm and Fourier laws also hold on statistical
average, with position-independent electric and heat conductivities.

\begin{remark}
\textnormal{Here we compare the values of the average electric and heat conductivities for the three
considered situations. Indeed, looking at the expressions \eref{eq
  Ge}--\eref{eq Gh}, one sees that they only differ from a multiplicative
factor, which depends on the nature of the particles. In the electric
and heat cases, we found the
following order: BE $>$ MB $>$ FD, for all $x_0 > 0$ (the common admissible region).}
\end{remark}

\subsection{Random Matrix Theory}

As mentioned in the introduction, the classical EY-model describing particle and energy transport (\Fig~\ref{The Classical System}) has the following
important property: it is chaotic. More precisely, the cells, \emph{without}
the discs, which are making up the system are ergodic. In analogy, we assume the following.
\begin{assumption}
The quantum dots are classically chaotic.
\end{assumption}
Under this assumption, it is natural to describe the transport properties of the quantum
dots with random matrix
theory (RMT) \cite{Beenakker}. More precisely, we will consider that the
quantum dots are described by the Wigner-Dyson circular
ensembles (with $\beta = 1, 2$).

Here, we restrict our attention to the one channel situation,
\ie~we assume that $M_i = 1$ for $i \in \{\L, \R, 1, \dots,
N\}$. Now we consider that the $3 \times 3$
complex random matrices, $S^{(1)},\dots,S^{(N)}$, are independent and identically distributed
over $U(3)$ with some measure. More specifically, we consider
the following well-known cases:

\begin{itemize}
\item[(1)] The Circular Orthogonal Ensemble (COE, $\beta = 1$)
  describes the case of time-reversible systems.
\item[(2)] The Circular Unitary Ensemble (CUE, $\beta = 2$) corresponds to the case
  in which the time-reversal symmetry is broken. Such a situation appears,
  for example, when the system is submitted to some  external magnetic field.
\end{itemize}
 In the sequel, we shall be interested mainly in the following
quantities: 
\begin{itemize}
\item[(i)] The universal profile $A_1, \dots, A_N$, which gives the
  shape of the temperature and chemical potential
profiles. In the subsequent plots, we naturally set $A_0 = A_\L = 0$ and $A_{N+1}=A_\R = 1$.
\item[(ii)] The universal conductivity $\kappa(N)$, which gives the electric
  and heat currents through the chain of $N$ quantum dots.
\end{itemize}
As explained previously, we shall compare the classical and quantum
situations by working in terms of the probability matrices,
$P^{(1)},\dots,P^{(N)}$, and scattering matrices,
$S^{(1)},\dots,S^{(N)}$, respectively (see \Remark~\ref{remark on
  B}). 

For a review on RMT we refer to \cite{Beenakker}. To implement the numerical simulations,
we have followed \cite{Mezzadri}. In the sequel, the averages
are made over an ensemble of $10^4$-$10^6$ realizations, depending on
the size $N$ of the system.

\subsection{The Global Transmission Probabilities}\label{The S-matrix circular ensembles}

We denote by $\langle  t^{(k)}_{ij}  \rangle_{1}$ and $\langle 
t^{(k)}_{ij}  \rangle_{2}$ the average of the transmission
probabilities in COE\break ($\beta = 1$) and CUE ($\beta = 2$),
respectively. It turns out that one can determine their exact values \cite{Beenakker}:
\begin{equation}\label{Eq Transmission RMT}
\langle t^{(k)}_{ij}  \rangle_{1} = \left\{\begin{array}{ll}
1/4 & \mbox{if} \hspace{3mm}  i \not = j\\
1/2 & \mbox{if} \hspace{3mm}  i =
j\end{array}\right. \hspace{5mm} \mbox{and} \hspace{5mm} \langle 
t^{(k)}_{ij}  \rangle_{2} = \frac{1}{3}~, \hspace{3mm} \forall i,j~.
\end{equation}
Now we use the construction explained in Appendix~H to build the
scattering matrix $S$ associated to $N$ quantum dots. As the reader may guess, it is then very difficult to make any analytical statement
  concerning the averages $\langle t_{ij}  \rangle_{1}$ and
  $\langle  t_{ij}  \rangle_{2}$. We thus turned to numerical simulations,
  and made the following observations for large system sizes $N$:
\begin{itemize}
\item[(1)] The classical and quantum global transmission probabilities, $t^{\rm{cl}}_{ij}$ and $t^{\rm{qu}}_{ij}$, are
  \emph{different}, but their averages, $\langle  t^{\rm{cl}}_{ij}
  \rangle_{\beta}$ and $\langle  t^{\rm{qu}}_{ij}
  \rangle_{\beta}$, are the
  same, showing that \emph{on
  average} the interferences are negligible when $N$ is large. 
\item[(2)] The averages $\langle  t_{ij}  \rangle_{\beta}$ do not depend on the system size $N$.
\item[(3)] They are symmetric:
$\langle  t_{ij}  \rangle_\beta = \langle  t_{ji} 
\rangle_\beta$. 
\item[(4)] The average couplings between the terminals, $\langle  t_{ij}  \rangle_{\beta}$, depend on
  $|i-j|$ and are \emph{short ranged}, \ie $\langle t_{i j} \rangle_{\beta} \simeq 0$ if $|i-j|
> 2$. In particular, the average probability matrix has \emph{essentially} the following form:
\begin{equation*}
\langle  P  \rangle_{\beta} = \matrice{* & * & * &  & & & &  \\  \hspace{0.001mm} * & * & * & &  & & &\\ &
  * & * & *  & & &  & \\ &  & * & * & * & & & \\ & & & \ddots & \ddots
  & \ddots & &
  \\ & & & & * & * & * &  \\  & & & & & * & * & *  \\
  & & & &  &  * & * & *}~.
\begin{picture}(0,0)
\put(-62,37){\Huge{0}}
\put(-128,-47){\Huge{0}}
\end{picture}
\end{equation*}
The values of the matrix elements represented by 0 are smaller than 0.05.
\end{itemize}

\begin{figure}[htbp]
\begin{center}
\centerline{\hspace{3mm}\psfig{file=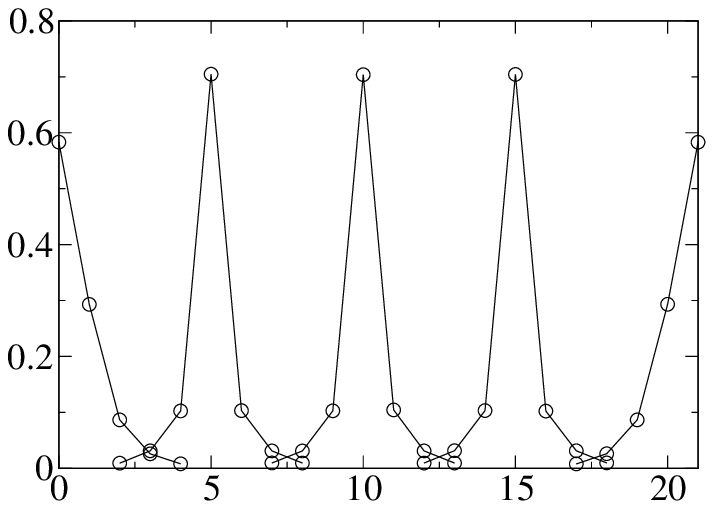,height=8cm,angle=-90}
  \psfig{file=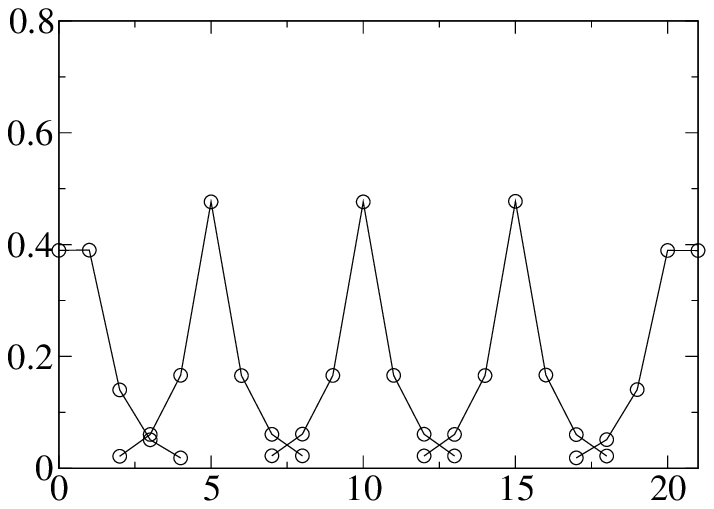,height=8cm,angle=-90}}
\caption{The (classical and quantum) average transmission probabilities, $\langle  t_{i\ell} 
\rangle_{\beta}$, for $\ell \in \{\L, 5, 10, 15, \R\}$ (with $\L = 0$ and
$\R = 21$) and some $i \in \{\ell - 4, \dots, \ell + 4\}$. For
example, the middle curves give: $\langle  t_{7, 10} 
\rangle_{\beta}$, $\dots$, $\langle  t_{10, 10} 
\rangle_{\beta}$, $\dots$, $\langle  t_{13, 10} 
\rangle_{\beta}$. Left: COE ($\beta
= 1$). Right: CUE ($\beta
= 2$).}\label{Fig Smatrix}
\end{center}
\begin{picture}(0,0)
\put(155,235){COE}
\put(385,235){CUE}
\end{picture}\vspace{-15mm}
\end{figure}

A typical example illustrating the above observations is given in
\Fig~\ref{Fig Smatrix}, where $N = 20$. In this figure, five terminals are considered: $\ell \in \{\L, 5, 10, 15, \R\}$ (with $\L = 0$ and
$\R = 21$). The different curves correspond to $\langle  t_{i\ell} 
\rangle_{\beta}$ for some $i \in \{\ell - 4, \dots, \ell + 4\}$. For
example, the middle curves give the following values: $\langle  t_{7, 10} 
\rangle_{\beta}$, $\dots$, $\langle  t_{10, 10} 
\rangle_{\beta}$, $\dots$, $\langle  t_{13, 10} 
\rangle_{\beta}$.

\vspace{-3mm}


\subsection{The Profiles and Currents}\label{The profiles and conductances}

Let us recall the expressions for the coefficient $A_i$ and for the
universal conductivity $\kappa(N)$:
\begin{equation*}
A_i = - \sum_{j = 1}^{N} (\Gamma_{\rm C}^{-1})_{ij} \ t_{j \R}
\hspace{5mm} \mbox{and} \hspace{5mm} \kappa(N) = N \left[t_{\L\R} +
  \sum_{i = 1}^{N} \sum_{j = 1}^{N} t_{\L j} \ (\Gamma_{\rm
  C}^{-1})_{ji} \ t_{i \R}\right]~.
\end{equation*}

\begin{figure}[htbp]
\begin{center}
\centerline{\hspace{5mm}\psfig{file=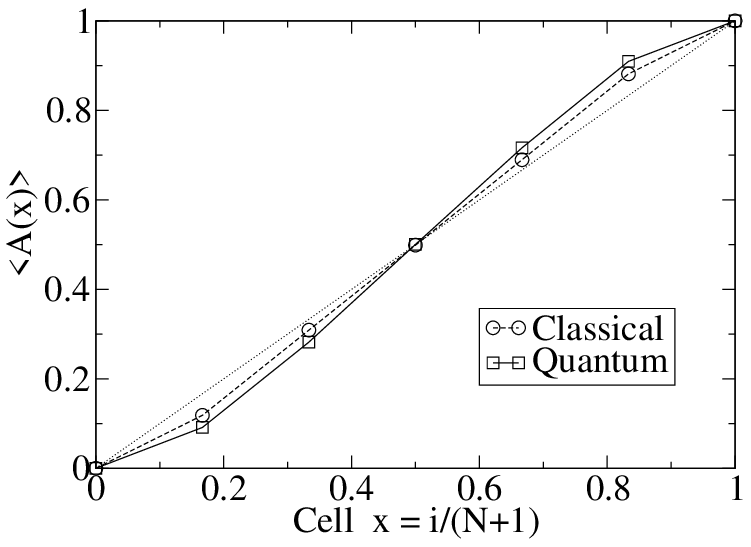,height=8cm,angle=-90} \psfig{file=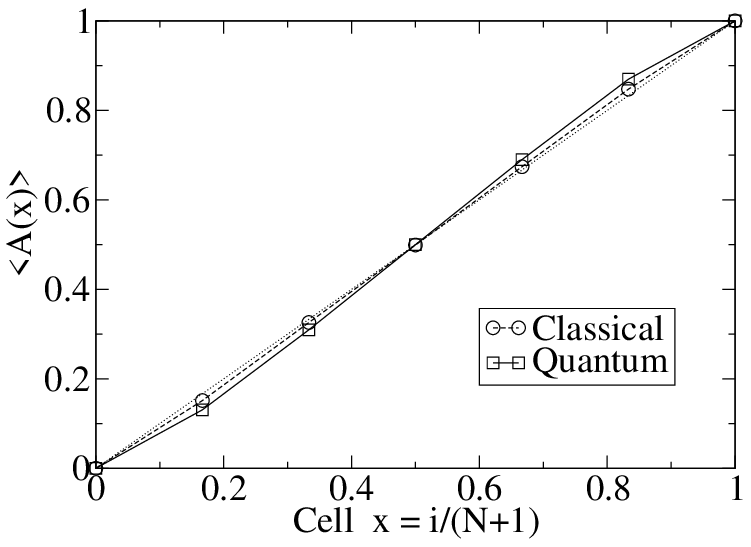,height=8cm,angle=-90}}
\caption{The classical and quantum average profiles $\langle
A(x) \rangle_\beta$ for $N =
  5$. The diagonal lines are added to guide the eye. Left: COE ($\beta
= 1$). Right:
  CUE ($\beta
= 2$).}\label{Fig B profile}
\end{center}
\begin{picture}(0,0)
\put(30,205){COE}
\put(262,205){CUE}
\end{picture}\vspace{-15mm}
\end{figure}

In \Fig~\ref{Fig B profile}, we show the average profiles $\langle
A(x) \rangle_\beta$ for $N =
5$. If $T_\L < T_\R$, these curves correspond essentially to the temperature
profiles along the system. (The same holds for the chemical potential profiles if
$\mu_\L < \mu_\R$). We
see that all these profiles are not linear, the non-linearity being more
stressed in the quantum situation, showing the effect of the
interferences. However, we found that all these profiles, while conserving their shape, get closer and closer to linear as the number of dots increases  and that eventually they become linear in the limit $N \rightarrow \infty$.

\begin{figure}[h!]
\begin{center}
\centerline{\psfig{file=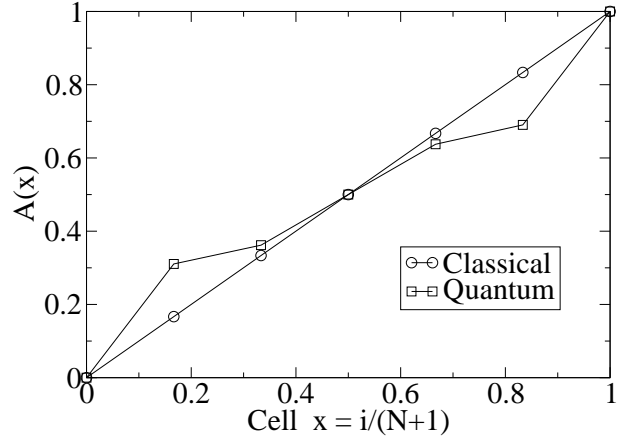,height=9cm,angle=-90}}
\caption{The profiles of the toy models in \Remark~\ref{rem toy} ($N =
  5$).}\label{Fig B profile Special2}
\end{center}\vspace{-9mm}
\end{figure}

\begin{remark}\label{rem toy}
\textnormal{In order to emphasise the role of interferences, let us
  consider the following simple toy models. We consider that the
  classical probability matrices are identical and given by
  $P^{(k)}_{ij} = 1/3$, for all $i,j$. In such a
  situation, the profile $A(x)$ is linear (for all $N$). On the other hand,
  the following unitary scattering matrices 
\begin{equation}
S^{(k)} = \frac{1}{\sqrt{3}} \matrice{ 1 & e^{\frac{2\pi}{3}i} &
  e^{\frac{2\pi}{3}i}\\ e^{\frac{2\pi}{3}i} & 1 & e^{\frac{2\pi}{3}i}
  \\ e^{\frac{2\pi}{3}i} & e^{\frac{2\pi}{3}i} & 1 }~,
\end{equation}
which satisfy the equiprobability property, $|S^{(k)}_{ij}|^2 =
\frac{1}{3}$, do not lead to a linear profile, showing the interference
effects (see \Fig~\ref{Fig B profile Special2}). A discussion about the conductivities in these cases is given in \Subsection~\ref{do not need chaos}.}
\end{remark}

\begin{remark}\label{remark toy model}
\textnormal{One may have noticed that the average temperature profiles differ from
  those in the EY-model. Indeed, in the  EY-model the
  temperature profiles were either linear,
  convex or concave functions of the cell
  position, depending on the values of the injection rates $\gamma_\L$ and $\gamma_R$. Basically, the reason was that, in the presence of a particle
  current, for example, going from right to left, the disc $i$ hears more often
  from cell $i+1$ than from cell $i-1$. It has therefore a greater
  tendency to equilibrate with the right than with the left, causing a
  curvature in the temperature profile. In particular, the temperature
  profile depends on all the boundary values $T_\L$, $T_\R$,
  $\gamma_\L$ and $\gamma_\R$. In the present model, the coupling among the terminals is governed by
  the scattering matrix $S$ and we have assumed it is
  energy-independent leading to a self-consistent temperature profile
  dependent only on $S$ and on the boundary values $T_\L$ and $T_\R$. However, we expect that in the
  general energy-dependent S-matrix situation, the temperature profile
  will also depend on $\mu_\L$ and $\mu_R$. To see whether in this
  case the
  EY-model and our model agree needs
  further investigations.} 
\end{remark}

We have shown in the preceding section that, if
$\mu_\L = \mu_\R$ and $T_\L < T_\R$, then the heat current $J_\L$ goes from
the hot reservoir (Right) to the cold one (Left), as one expects
from the second law of thermodynamics. In \Fig~\ref{Fig B profile},
one sees that the profiles are monotically increasing from left to right. This means that, on average, heat also flows locally from
hot to cold, \ie on statistical average, the second law of thermodynamics also holds
locally. 

All the preceding results concern statistical average values. In \Fig~\ref{Fig B profile
  Special}, we present different realizations and one sees that very different
profiles may occur. Roughly speaking, by choosing an appropriate combination of
the S-matrices, $S^{(1)}, \dots, S^{(N)}$, one may generate almost any profile.

\begin{figure}[h!]
\begin{center}
\centerline{\psfig{file=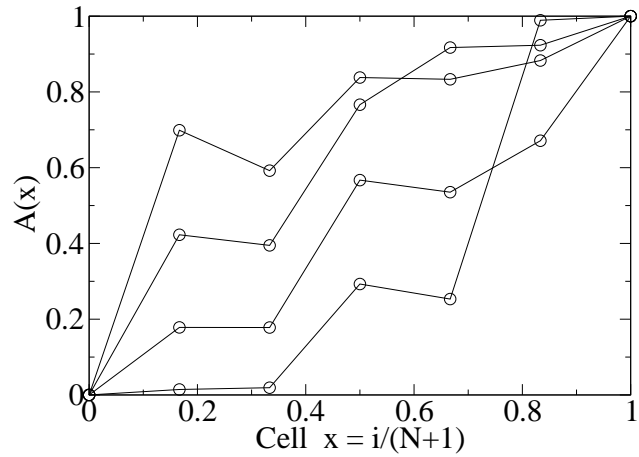,height=9cm,angle=-90}}
\caption{The quantum profiles for some realizations ($N =
  5$).}\label{Fig B profile Special}
\end{center}
\end{figure}

In particular, one sees in \Fig~\ref{Fig B profile
  Special} that there exist \emph{particular} realizations for which
  the heat current flows \emph{locally} from cold to hot. In other words, the profiles are not monotone increasing
for every realization, but only on statistical average. Such phenomena are well
known in mesoscopic systems (see \eg~\cite{Buttiker-Symmetry,ButtikerNR,ButtikerPotential}). 

We observed that the quantum dots about which such phenomena
occur (\ie such that $A_{i} < A_{i-1}$
and consequently $\kappa_{\rm e}(i-1) < 0$ and  $\kappa_{\rm h}(i-1) < 0$) have typically a very high
reflection probability $t_{ii}$. As the system size $N$ grows, we see
that such phenomena diminish in intensity and frequency. Basically, we
expect that, when the system
 is macroscopic, the second law of thermodynamics also holds locally in
 every realization. 

Moreover, in the thermodynamic limit $N \rightarrow \infty$, we see that the
average profiles become linear in all situations (COE, CUE, classical and
quantum):
\begin{equation}\label{Profile Ax}
\langle A(x) \rangle_\beta = x~, \hspace{5mm} \forall x \in [0,1]~. 
\end{equation}
Consequently,
\begin{equation}
\langle V(x) \rangle_\beta = V_\L + (V_\R - V_\L) \ x \hspace{5mm} \mbox{and}  \hspace{5mm}
\langle T(x) \rangle_\beta = T_\L + (T_\R - T_\L) \ x~.
\end{equation}

\begin{figure}[h!]
\begin{center}
\psfig{file=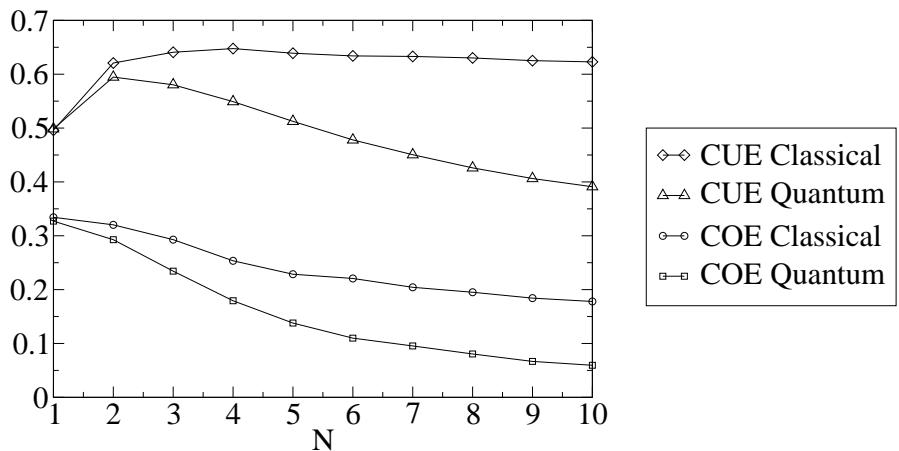,height=13cm,angle=-90}
\caption{The average universal conductivities $\langle \kappa(N)
  \rangle_\beta$ for small $N$.}\label{Fig Sigma Small N}
\end{center}
\end{figure}

We next present some results concerning the universal conductivity
$\kappa(N)$. In \Fig~\ref{Fig Sigma Small N}, we show $\langle \kappa(N)
\rangle_\beta$ for small $N$. Observe that in the COE cases the conductivity decreases with $N$ while in the CUE cases it first increases and only then decreases. 

\newpage

\begin{figure}[h!]
\begin{center}
\centerline{\hspace{3mm}\psfig{file=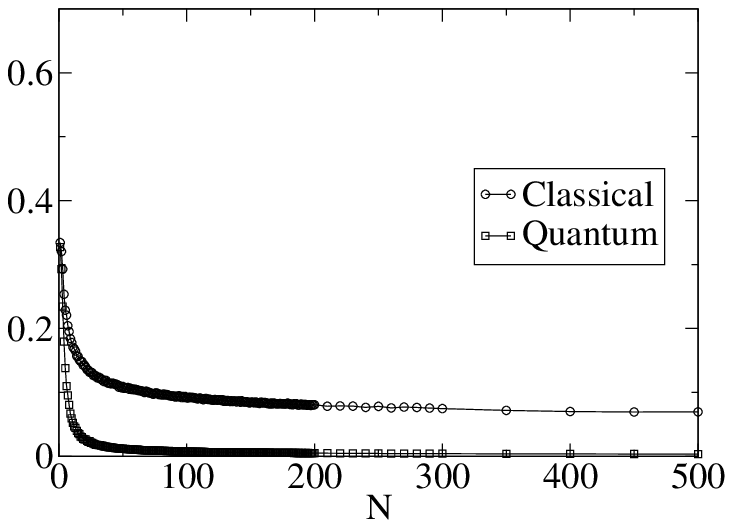,height=8cm,angle=-90} \psfig{file=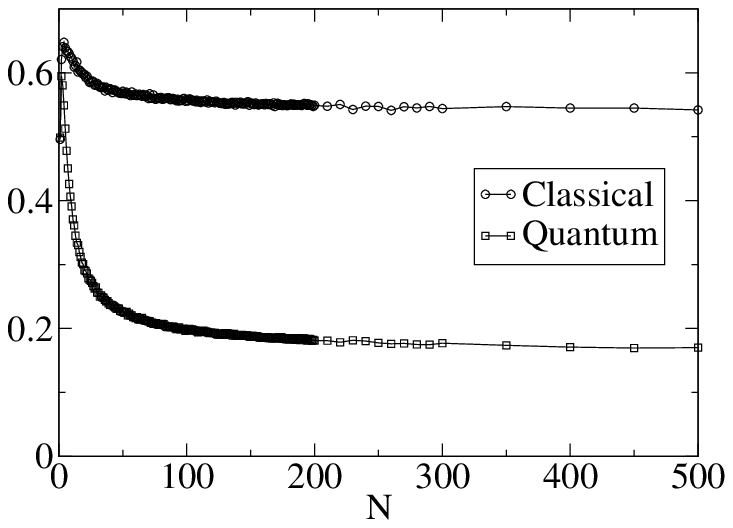,height=8cm,angle=-90}}
\caption{The average universal conductivities $\langle \kappa(N)
  \rangle_\beta$ as a function of $N$.}\label{Fig Sigma}
\end{center}
\begin{picture}(0,0)
\put(158,195){COE}
\put(390,195){CUE}
\end{picture}\vspace{-5mm}
\end{figure}

For larger values of $N$, we see in \Fig~\ref{Fig Sigma} that the average universal conductivity seems to admit a finite limit $\langle \kappa^{\infty} \rangle_\beta$ as $N
\rightarrow \infty$. To support this observation and obtain a numerical
value for $\langle \kappa^{\infty} \rangle_\beta$, we have proceeded as
follows. We found that the curves in \Fig~\ref{Fig Sigma} are very
well fitted (the largest chi-square being 0.035) by
functions (not shown in the figures) of the following form: $\langle  \kappa^{\infty} 
\rangle_\beta + c_\beta/N^\alpha$. Here, $c_\beta$ is some positive
constant and $\alpha$ is an exponent found to be the same in COE and
CUE, and given by $\alpha^{\rm cl} = 1/2$ and $\alpha^{\rm qu} = 1$. Moreover,
\begin{eqnarray*}
\mbox{Classical:} & &  \hspace{10mm}\langle \kappa^{\infty} \rangle_1 = 0.057  \hspace{7mm} \mbox{and}
\hspace{7mm} \langle  \kappa^{\infty}  \rangle_2 = 0.531~.\\ 
\mbox{Quantum:} & & \hspace{10mm} \langle  \kappa^{\infty} \rangle_1 = 0.005  \hspace{7mm} \mbox{and}
\hspace{7mm} \langle  \kappa^{\infty}  \rangle_2 = 0.164~.
\end{eqnarray*}

Let us denote by $\langle \kappa^{\infty}_{\rm e} \rangle_\beta$ and
$\langle \kappa^{\infty}_{\rm h} \rangle_\beta$ the corresponding limits:
\begin{equation}\label{exp Ge}
\langle \kappa^{\infty}_{\rm e} \rangle_\beta = \frac{a}{\Sigma} \ \frac{e^2 C(0)}{h} \ \langle 
\kappa^{\infty}  \rangle_\beta
\end{equation}
and
\begin{equation}\label{exp Gh}
\langle \kappa^{\infty}_{\rm h} \rangle_\beta  = \frac{a}{\Sigma} \ \frac{k_\B^2 T}{h} \ \frac{C(0) C(2) - C(1)^2}{C(0)} \
\langle  \kappa^{\infty}  \rangle_\beta~.
\end{equation}
These numerical results show that the global Ohm and Fourier laws hold on
\emph{statistical average} in our chain of quantum dots with self-consistent reservoirs. Since the average temperature and chemical potential profiles 
become linear in the limit $N \rightarrow \infty$, it follows that the local Ohm and Fourier laws also hold on
statistical average, with position-independent electric and heat conductivities, 
\ie 
$$
\langle \mathcal{I}(x) \rangle_\beta = - \langle \kappa^{\infty}_{\rm
  e}(x) \rangle _\beta \, \nabla \langle V(x) \rangle_\beta
\hspace{4mm}\mbox{and} \hspace{4mm} \langle \mathcal{J}(x)
\rangle_\beta = - \langle \kappa^{\infty}_{\rm h}(x) \rangle_\beta \, \nabla \langle T(x) \rangle_\beta ~,
$$ 
with $\langle \kappa^{\infty}_{\rm e}(x)  \rangle_\beta = \langle \kappa^{\infty}_{\rm e} \rangle_\beta$ and $\langle \kappa^{\infty}_{\rm h}(x) \rangle_\beta = \langle \kappa^{\infty}_{\rm h} \rangle_\beta$ for all $x \in (0,1)$. Here we have kept the global gradients \eref{global gradients} constant while taking the limit $N \rightarrow \infty$.

\begin{remark}\label{rem optimal}
\textnormal{In the single channel case ($M_i = 1$, for all $i$), the \emph{optimal}
  bounds \eref{bounds} become $0 <
  \kappa(N) < N$.  Let us consider the two extreme situations in which all
  the quantum dots are identical.
\begin{itemize}
\item[(1)] Total back-reflection:
$$
S^{(k)} = \matrice{ 1 & 0 &
  0\\ 0 & 1 & 0  \\ 0 & 0 & 1 }~.
$$
In this case, we found $\kappa(N) = 0$ for all $N$.
\item[(2)] Optimal
transmission:
$$
S^{(k)} = \matrice{ 0 & 0 &
  1\\ 0 & 1 & 0  \\ 1 & 0 & 0 }~.
$$
This corresponds to the case in which all the intermediate terminals are
  disconnected from the system and the transmission through the system
  is optimal (\ie $t_{\L \R} = t_{\R \L} =
  t_{11} = \dots = t_{NN} =  1$ and all other $t_{ij} = 0$). We found $\kappa(N) = N$, for all $N$. Hence, a perfect
  conductor has infinite electric and heat conductivities in the limit $N
  \rightarrow \infty$.
\end{itemize}}
\end{remark}

Although we have shown that typically (in the
  sense of RMT) $\kappa(N)$ converges to a finite limit as $N \rightarrow
  \infty$, one deduces from the examples in \Remark~\ref{rem optimal} that the linear geometry, represented in \Fig~\ref{The Quantum System}, is not sufficient to
  ensure the existence of such a limit. The characterization of the sequences
  of scattering matrices, $\left(S^{(k)}\right)_{k = 1}^{\infty}$, for which
  the limit, $\lim_{N \rightarrow \infty} \kappa(N)$, exists and is
  finite remains an open problem. Note however that, approximate results were obtained in some particular cases \cite{Amato,roy}. Note also that the particular ordered cases, in which all the local scattering matrices are identical,  are not simpler to handle with the relations given in Appendix~H.
  
In \Fig~\ref{Fig Sigma}, one sees two manifestations of weak
localization: 
\begin{itemize}
\item[(1)] The classical conductivity is larger than the
quantum conductivity. Hence, the effect of the interferences is to
decrease the intensity of the electric and heat currents. Note that this effect
is rather subtle. Indeed, one has
\begin{equation}
\langle \kappa(N)  \rangle_\beta = N \left[\langle  t_{\L\R}  \rangle_\beta
  +
  \sum_{i,j = 1}^{N} \langle  t_{\L j} \ 
  (\Gamma_{\rm C}^{-1})_{ji} \ t_{i \R}   \rangle_\beta\right]~.
\end{equation}
Since the average global transmission
probabilities are the same in the classical and quantum situations,
the effect of the interferences on the conductivity must be due to
correlations among the random variables $t_{ij}$. 
\item[(2)] The application of an external field (CUE) increases the values
of the conductivities and consequently increases the intensity of
the electric and heat currents. This can be understood as
follows: In equation~\eref{Eq Transmission RMT} and in \Fig~\ref{Fig
  Smatrix}, one sees that the presence of a magnetic field will (on
average) increase the transmissions probabilities. 
\end{itemize}

\begin{remark}\label{Remark Amato}
\textnormal{The case of electronic transport at low temperature was
  already considered by D'Amato and Pastawski \cite{Amato}. Modelling
  the system as a nearest-neighbour tight-binding Hamiltonian they
  found \emph{approximate} expressions for the global transmission probabilities
  $t_{ij}$, which turn out to satisfy the properties (2)--(4) presented in \Subsection~\ref{The S-matrix
  circular ensembles}. From these approximate $t_{ij}$ they deduced that, in the
  limit the system size $N \rightarrow \infty$, the
  self-consistent chemical potential profile is linear and the
  universal conductivity is finite, in agreement with our results. However, in our opinion,  these (nice) results need further analysis to constitute a rigorous derivation of Ohm's law in their model.}
\end{remark}

\subsection{Chaos, Disorder and Decoherence}\label{do not need chaos}

\begin{figure}[h!]
\begin{center}
\centerline{\psfig{file=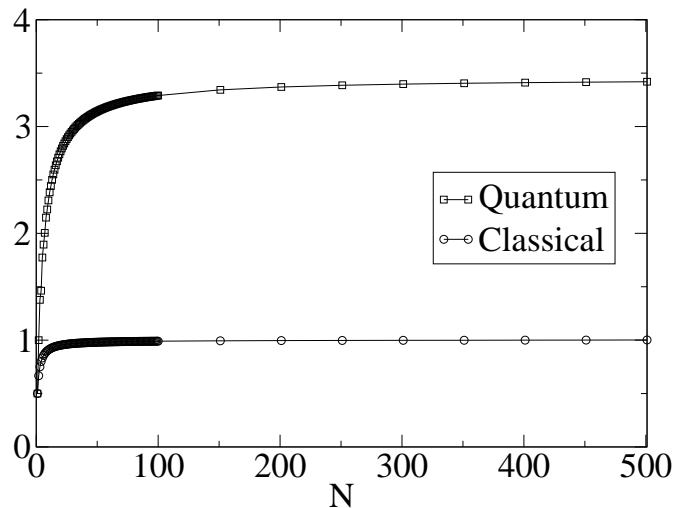,height=10.5cm,angle=-90}}
\caption{The universal conductivities $\kappa(N)$  of the toy models in \Remark~\ref{rem toy}.}
\label{Conductivity toy model}
\end{center}
\end{figure}

In this paper,  we were not interested in finding the optimal conditions to obtain Ohm and Fourier laws, but rather to establish a quantum version of the classical chaotic EY-model. This led us naturally to use RMT to characterize the quantum dots. Nevertheless, one may wonder whether chaoticity or randomness, as modelled by RMT, is essential to obtain normal transport, \ie for Ohm and Fourier laws to hold. The answer is no, since, for example, the two toy models presented in \Remark~\ref{rem toy}, which may be interpreted as ordered non-chaotic cases, do lead to Ohm and Fourier laws. Indeed, one sees in \Fig~\ref{Conductivity toy model} that their corresponding universal conductivities converge to finite values as the number of dots increases. Observe however that, contrary to the RMT situations, the conductivities increase with the number of dots and the "quantum" universal conductivity is higher than the "classical" one. This shows in particular that, although interferences are typically (in the sense of RMT) destructive, they may be constructive in some particular cases. Actually, this phenomenon is typical in \emph{ordered} chains (see below).

\newpage

\begin{figure}[htbp]
\begin{center}
\centerline{\hspace{5mm}\psfig{file=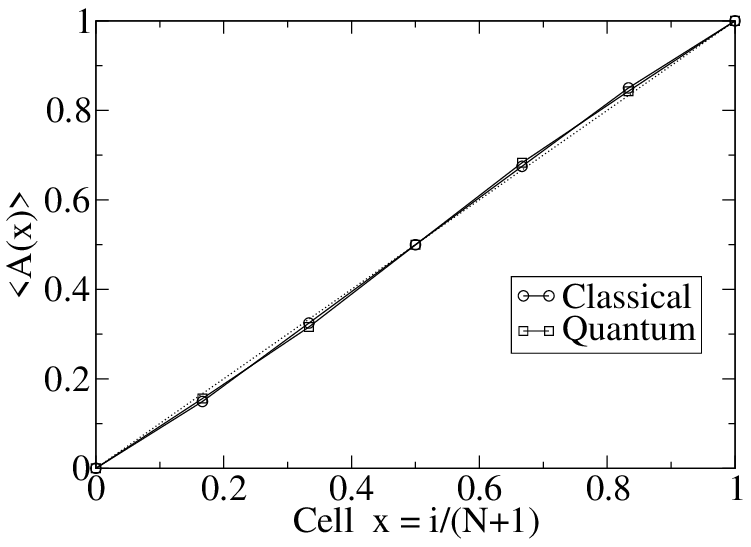,height=8cm,angle=-90} \psfig{file=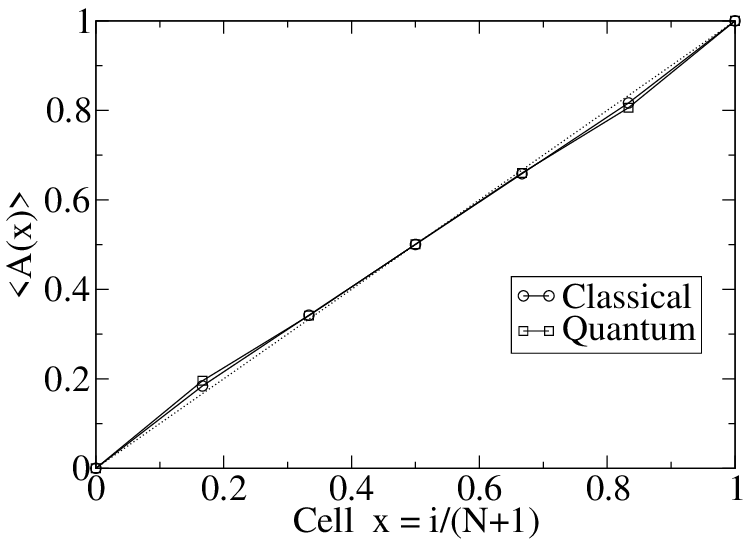,height=8cm,angle=-90}}
\caption{The average profiles $\langle
A(x) \rangle_\beta$ in the ordered cases ($N =
  5$). The diagonal lines are added to guide the eye. Left: COE ($\beta
= 1$). Right:
  CUE ($\beta
= 2$).}\label{Fig B profile ordered}
\end{center}
\begin{picture}(0,0)
\put(30,205){COE}
\put(262,205){CUE}
\end{picture}\vspace{-20mm}
\end{figure}

\begin{figure}[h!]
\begin{center}
\centerline{\psfig{file=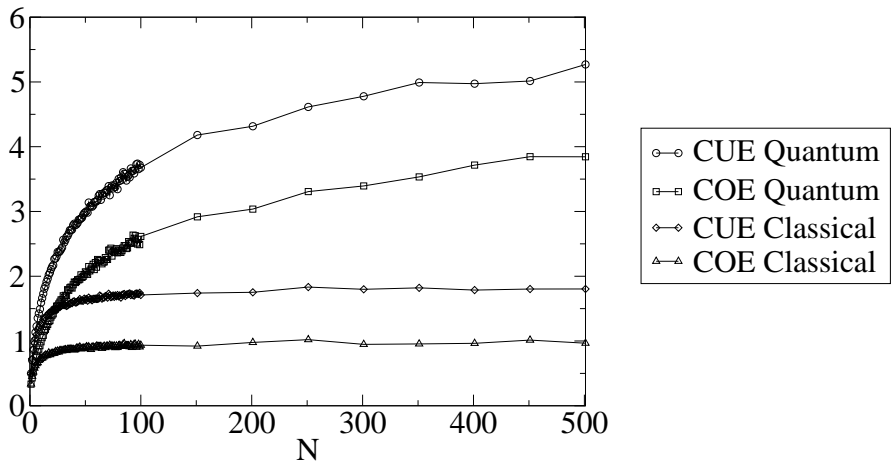,height=13cm,angle=-90}}
\caption{The average universal conductivities $\langle \kappa(N)
  \rangle_\beta$ in the ordered chains.}
\label{Conductivity Ordered model}
\end{center}\vspace{-5mm}
\end{figure}

Another natural investigation concerns the effects of disorder. We say that the chain is \emph{ordered} if all the local scattering matrices are identical (\ie $S^{(1)} = \dots = S^{(N)}$) and \emph{disordered} otherwise. The results presented in \Figs~\ref{Fig B profile} and \ref{Fig Sigma} concern the disordered chaotic cases and in \Figs~\ref{Fig B profile Special2} and \ref{Conductivity toy model} one can find two examples of ordered non-chaotic cases. The ordered chaotic situation is obtained by using RMT with the condition that for each realization $S^{(1)} = \dots = S^{(N)}$. The results are given in \Figs~\ref{Fig B profile ordered} and \ref{Conductivity Ordered model}. Observe that by applying a magnetic field the "s-shapes" of the profiles are reversed and that the conductivities behave qualitatively as in the (ordered) toy model cases.  Note also that in the quantum COE situation, the asymptotic value of the universal conductivity is about three order of magnitude larger in the ordered case than in the disorder one. This shows that disorder may have a very strong localization effect.

\newpage

Finally, let us discuss the effects of decoherence. The idea is to view the self-consistent reservoirs as an effective environment acting on a chain of coherent quantum dots. For simplicity, let us assume that all the self-consistent reservoirs are equally coupled to the dots, \ie $1 - |S^{(k)}_{22}|^2 = \lambda \in [0,1]$ for all $k$. (In the simulations, the values of the coupling parameter $\lambda$ are taken in small intervals of width $0.01$.) Then, $\lambda = 0$ corresponds to the completely decoupled situation and $\lambda = 1$ corresponds to the maximally coupled one.  

\begin{figure}[h!]
\begin{center}
\centerline{\psfig{file=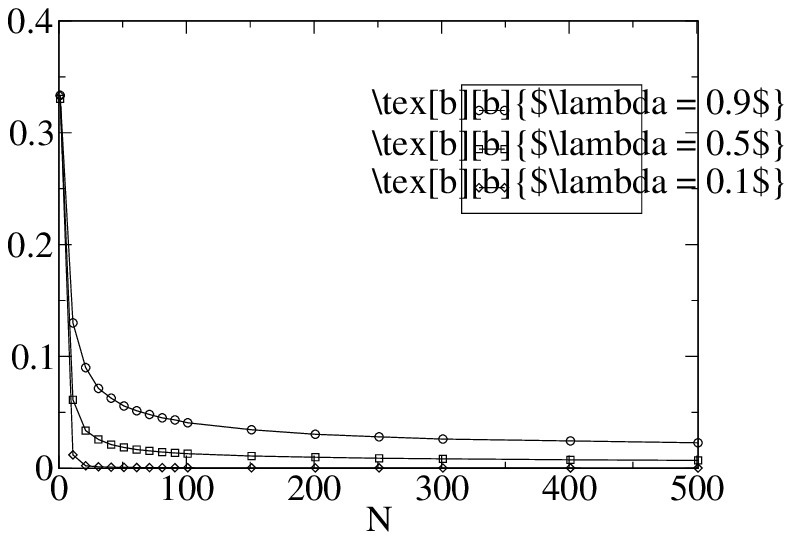,height=10cm,angle=-90}}
\caption{The average universal conductivity $\langle \kappa(N)
  \rangle_1$ in the quantum disordered chain, for three different values of the coupling parameter $\lambda$.}
\label{Conductivity Ordered model Coupling}
\end{center}
\begin{picture}(0,0)
\put(125,235){COE}
\end{picture}\vspace{-5mm}
\end{figure}

For any fixed value of $\lambda$, we found that the profiles and conductivities are qualitatively similar to those obtained previously. Interestingly, when $\lambda$ increases, we observed that the conductivity increases in the disordered cases while it decreases in the ordered ones. As an illustration, we show in \Fig~\ref{Conductivity Ordered model Coupling} the results for three different values of $\lambda$ in the quantum disordered chaotic chain. In this case, the profiles (not shown) are similar to those obtained in \Fig~\ref{Fig B profile} and tend to get closer and closer to linear as the coupling parameter $\lambda$ increases. 

\section{Concluding Remarks}

We have presented a model for charge and heat transport using the Landauer-B\"uttiker scattering approach. In the
linear response regime, we have seen that the Onsager relations
hold. Then, assuming that the scattering
matrix $S$ does not
depend on the energy, we have shown that the transport matrix $L$ is real positive
semi-definite and we have characterized the equilibrium and non-equilibrium  
states of the system in terms of the currents and in terms of the
entropy production. In particular, we
have shown that our multi-terminal system satisfies the first and second laws of
thermodynamics. 

In order to obtain an effective quantum version of the Eckmann-Young model with
rotating discs, we have imposed the self-consistency
condition: $I_i = J_i = 0$ for $i=1,\dots,N$. This condition led
to expressions for the temperature and chemical potential profiles, which turn out to be
independent of the nature of the particles, as
well as to expressions for the electric and heat currents
going through the system. Finally, we have presented some numerical
results, using random matrix theory, supporting the validity of Ohm
and Fourier laws in our model. 

Let us point out that we could handle the three physical
cases, Maxwell-Boltzmann, Fermi-Dirac and Bose-Einstein, on
the same footing because they share the following two features: (i) the specific form \eref{form g} of $f$ and (ii) the ratio
$\mathcal{R}$, defined in \eref{Ratio R}, satisfies $0 < \mathcal{R} <
1$. Basically, we expect that any
distribution function $f$ satisfying (i)-(ii) will lead to the same results. 

For further investigations, it would be interesting to study the
general energy-dependent S-matrix situation and to compare the corresponding
predictions (under RMT) with those of the EY-model. One may also analyse the nonlinear transport
properties through the full counting statistics (FCS)
\cite{Levitov-Lesovik,levitov2,Bagrets,Pilgram-Jordan,Pilgram2,Kindermann,Pilgram,Saito1}. More mathematically oriented investigations would be to
obtain some rigorous results concerning the global composite
scattering matrix $S$  and consequently to prove the
existence of the finite limit, $\lim_{N \rightarrow
  \infty}\kappa(N)$, which ensures the validity of Ohm and Fourier laws in
our model.

\section*{Acknowledgements}

The author has benefited from numerous helpful discussions with M. B\"uttiker
and his group members, H. Foerster, A. M. Lunde, S. Nigg, C. Petitjean,
M. L. Polianski and J. Splettst\"osser, and also with
J.-P. Eckmann, M. Hairer, J. Jacquet, Ph. Jacquod, C. Mej\'ia-Monasterio,
M. Moskalets, C.-A. Pillet, L. Rey-Bellet, E. Sukhorukov, P. Wittwer and C. Zbinden, to all of whom he wishes to
express his sincere gratitude. This work was partially supported by
the Swiss National Science Foundation and the Erwin Schr\"odinger International Institute for Mathematical Physics.

\renewcommand{\theequation}{A.\arabic{equation}}
\setcounter{equation}{0} 
\section*{Appendix A: The Coefficients $\mathbf{C(n)}$}\label{The coefficients}

Here we derive some properties of the
coefficients $C(0)$, $C(1)$ and $C(2)$. For this, let $n = 0$, $1$ or
$2$. Then\\ \\
(i) If $x_0 \in (-1,\infty)$, then $C^{\rm MB}(n)
> 0$. Indeed, this follows from 
\begin{equation}\label{expl MB}
 C^{\rm MB}(0) = e^{-x_0}~, \hspace{2mm} C^{\rm MB}(1) = (1+x_0)
e^{-x_0}~, \hspace{2mm}C^{\rm MB}(2) = (2+2 x_0+x_0^2~)
e^{-x_0}~.
\end{equation}
(ii) If $x_0 \in (-\infty,\infty)$, one has $C^{\rm FD}(n) > 0$, and
if $x_0 \in (0,\infty)$, one has $C^{\rm BE}(n) > 0$. Indeed, if $x_0 >
0$, then clearly $C^{\rm FD}(n) > 0$ and $C^{\rm BE}(n) > 0$. For the Fermi-Dirac case, assume that $x_0 \leq 0$, then
\begin{eqnarray}
C^{\rm FD}(0) &=& \int_{x_0}^{\infty}
  \frac{ e^x}{(e^x + 1)^2} \ dx > 0~,\\
C^{\rm FD}(2) &=& \int_{x_0}^{\infty}
  \frac{x^2 e^x}{(e^x + 1)^2} \ dx > 0~.
\end{eqnarray}
Now,
\begin{equation}
C^{\rm FD}(1) = \int_{x_0}^{\infty}
  \frac{x e^x}{(e^x + 1)^2} \ dx~.
\end{equation}
Observe that $C^{\rm FD}(1) > 0$ if $x_0 = 0$ and
$C^{\rm FD}(1) \rightarrow 0$ as $x_0 \rightarrow -\infty$. Therefore,
the
property $C^{\rm FD}(1) > 0$, for all $x_0 \leq 0$, follows from 
\begin{equation}
\frac{d}{dx_0} C^{\rm FD}(1) = -\frac{x_0 e^{x_0}}{(e^{x_0} + 1)^2} \geq 0~.
\end{equation}

\renewcommand{\theequation}{B.\arabic{equation}}
\setcounter{equation}{0} 
\section*{Appendix B: The Ratio $\mathbf{\mathcal{R}}$}

Here, we show that for all $x_0$:
\begin{equation}\label{Lemma L0}
\mathcal{R} \equiv \frac{Q_1}{\sqrt{Q_2}} =
\frac{C(1)}{\sqrt{C(0) \cdot C(2)}} \in (0,1)~.
\end{equation}
Since $Q_1 > 0$ and $Q_2 > 0$, one has $\mathcal{R} > 0$, so
it only remains to show that $\mathcal{R} < 1$.\\ \\
(i) Let us first consider the Maxwell-Boltzmann case. Recalling the
explicit expressions \eref{expl MB}, one obtains ($x_0 \in (-1,\infty)$)
\begin{equation}
\mathcal{R}^2 = \frac{1 + 2 x_0 + x_0^2}{2 + 2 x_0 + x_0^2} < 1~.
\end{equation}
To show that $\mathcal{R} < 1$ in the other two cases, it is convenient to introduce the
real functional space
$L^2([a,\infty),dx)$ with scalar product $(\cdot, \cdot)_{a}$ and norm
  $\| \cdot \|_{a}$:
\begin{equation}
(f_1, f_2)_{a} = \int_{a}^{\infty} f_1(x) \ f_2(x) \ dx~,
  \hspace{10mm} \|
  f\|^2_{a} = (f, f)_{a}~.
\end{equation}
(ii) For the Bose-Einstein case, one has ($x_0 > 0$)
\begin{equation}
\mathcal{R} = \frac{\| \sqrt{x} \ g \|_{x_0}^2}{\|g \|_{x_0} \cdot
    \|x \ g \|_{x_0}}~, \hspace{5mm} \mbox{with} \hspace{5mm} g(x) = \frac{e^{x/2}}{|e^x - 1|}~.
\end{equation}
Hence, it is sufficient to show that 
\begin{equation}
\| \sqrt{x} \ g \|_{x_0}^2 < \|g \|_{x_0} \cdot \| x \ g
\|_{x_0}~.
\end{equation}
We have
\begin{equation}
\| \sqrt{x} \ g \|_{x_0}^2 = (\sqrt{x} \ g, \sqrt{x} \ g)_{x_0} = (g, x \ g)_{x_0}
< \| g \|_{x_0} \cdot \| x \ g \|_{x_0}~,
\end{equation}
where we have used the Cauchy-Schwarz inequality.
\newpage
\noindent
(iii) Finally, let us consider the Fermi-Dirac case. If $x_0 > 0$, then
we  proceed as in (ii). If $x_0 \leq  0$, then we observe that
\begin{equation}
\mathcal{R} \leq \frac{\| \sqrt{x} \ g \|_{0}^2}{\|g \|_{0} \cdot
    \|x \ g \|_{0}}~, \hspace{5mm} \mbox{with} \hspace{5mm} g(x) = \frac{e^{x/2}}{|e^x + 1|}~,
\end{equation}
and again proceed as in point (ii).

\renewcommand{\theequation}{C.\arabic{equation}}
\setcounter{equation}{0} 
\section*{Appendix C: The Entropy Production}\label{Appendix Reminder}

In this appendix, we derive the expression \eref{sigmas expression} for the
entropy production rate $\sigma_s$. However, let us first recall how a real positive
semi-definite matrix is defined. Let $\{e_j\} \subset \mathbb{R}^{2N+4}$ denote the
canonical basis in which $(e_i, L e_j) = L_{ij}$, where $(\cdot, \cdot)$
denotes the usual scalar product in $\mathbb{R}^{2N+4}$. Then the matrix
$L$ is said to be real positive
semi-definite if for any vector $V
\in \mathbb{R}^{2N+4}$ one has 
\begin{equation}
\sigma_s = (V, L V)  = \sum_{i} \sum_{j} L_{ij} V_i V_j \geq 0~.
\end{equation} 
Let us decompose the matrix elements $L_{ij}$ into a
symmetric and antisymmetric part:
\begin{equation}
L_{ij} = L^s_{ij} + L^a_{ij}~,
\end{equation}
where
\begin{equation}
L^s_{ij} = \frac{1}{2} (L_{ij} + L_{ji}) = L^s_{ji} \hspace{5mm} \mbox{and}
\hspace{5mm} L^a_{ij} = \frac{1}{2} (L_{ij} - L_{ji}) = -L^a_{ji}~.
\end{equation}
Then
\begin{equation}\label{Entropy}
\sigma_s = \sum_{i} \sum_{j} L^s_{ij} V_i V_j~.
\end{equation}
This shows that only the symmetric part of the transport matrix $L$ may contribute to the entropy
production. Observe that the symmetric transport coefficients,
$L^s_{ij}$, also satisfy the relations \eref{Property Lij} and
\eref{Property Lij2}. Since we shall only make use of these
properties, we see that we can assume that the matrix $L$ is symmetric.

The first idea is to decompose the sum \eref{Entropy} into four sums, one
associated to each block appearing in the matrix $L$. To simplify the notation, let us consider that the chain
contains $N-2$ quantum dots and then number as $1$ and $N$ the left and
right reservoirs, respectively. We shall then write $X_1, \dots, X_N$
for the first $N$
components of $V$ and $Y_1, \dots, Y_N$ for its last $N$
components. One may interpret this decomposition as follows: $X_i = \delta \mu_i/e$ and $Y_i = \delta
T_i / T$. Recalling that $L_{ij} = L_{ji}$, one has
\begin{equation}
\sigma_s  = \sum_{i=1}^{N} \sum_{j=1}^{N} L^{(0)}_{ij} X_i X_j +
\sum_{i=1}^{N} \sum_{j=1}^{N} (L^{(1)}_{ij} + L^{(1)}_{ji}) X_i Y_j +
\sum_{i=1}^{N} \sum_{j=1}^{N} L^{(2)}_{ij} Y_i Y_j~. 
\end{equation}
Now, using the relations
$L^{(1)}_{ij} = Q_1L^{(0)}_{ij}  = L_{ji}^{(1)}$ and $L^{(2)}_{ij} = Q_2
L^{(0)}_{ij}$, one obtains
\begin{equation}
\sigma_s = \sum_{i=1}^{N} \sum_{j=1}^{N} L^{(0)}_{ij} \left[X_i X_j +
2 Q_1 X_i Y_j + Q_2 Y_i Y_j\right]~.
\end{equation}
Let $Z_j = \sqrt{Q_2} Y_j$ and $\mathcal{R} =
Q_1/\sqrt{Q_2}$. Then
\begin{equation}
\sigma_s = \sum_{i=1}^{N} \sum_{j=1}^{N} L^{(0)}_{ij} \left[X_i X_j +
2 \mathcal{R} X_i Z_j + Z_i Z_j\right]~.
\end{equation}
Let us first consider the term:
\begin{equation}
\mathcal{T}_1 = \sum_{i=1}^{N} \sum_{j=1}^{N} L^{(0)}_{ij} X_i X_j~.
\end{equation}
We write
\begin{equation}
\mathcal{T}_1 = \sum_{i=1}^{N} L^{(0)}_{ii} X_i^2 + \sum_{i=1}^{N} \sum_{j=1}^{N}
L^{(0)}_{ij} X_i X_j (1 - \delta_{ij})~.
\end{equation}
Since $\sum_{j} L^{(0)}_{ij} = 0$, one has $L^{(0)}_{ii} = - \sum_{j}
L^{(0)}_{ij} (1-\delta_{ij})$ and therefore
\begin{equation}
\mathcal{T}_1 = -\sum_{i=1}^{N} \sum_{j=1}^{N} L^{(0)}_{ij} X_i^2 (1-\delta_{ij})+ \sum_{i=1}^{N} \sum_{j=1}^{N}
L^{(0)}_{ij} X_i X_j (1 - \delta_{ij})~.
\end{equation}
Now comes a trick:
\begin{equation}
\sum_{i=1}^{N} \sum_{j=1}^{N} L^{(0)}_{ij} X_i^2 (1-\delta_{ij}) = \sum_{i=1}^{N} \sum_{j=1}^{N} L^{(0)}_{ij} X_j^2 (1-\delta_{ij})~.
\end{equation}
Hence 
\begin{equation}
\mathcal{T}_1 = -\sum_{\scriptsize\begin{array}{c}
    i,j = 1\\ i < j \end{array}}^{N} L^{(0)}_{ij} \underbrace{(X_i^2
    -2 X_i X_j + X_j^2)}_{(X_i- X_j)^2}~.\label{Equ XX}
\end{equation}
This relation is well known (see \eg \cite{Buttiker-Symmetry}). Next we consider the second term:
\begin{equation}
\mathcal{T}_2 = 2 \mathcal{R} \sum_{i=1}^{N} \sum_{j=1}^{N} L^{(0)}_{ij} X_i Z_j
= 2 \mathcal{R} \sum_{i=1}^{N} L^{(0)}_{ii} X_i Z_i + 2 \mathcal{R} \sum_{i=1}^{N} \sum_{j=1}^{N}
L^{(0)}_{ij} X_i Z_j (1 - \delta_{ij})~.
\end{equation}
Using the relation $L^{(0)}_{ii} = - \sum_{j}
L^{(0)}_{ij} (1-\delta_{ij})$ and the identities
\begin{equation}
\sum_{i=1}^{N} \sum_{j=1}^{N} L^{(0)}_{ij} X_i Z_i (1 -
\delta_{ij}) = \sum_{i=1}^{N} \sum_{j=1}^{N} L^{(0)}_{ij} X_j Z_j (1 - \delta_{ij})~,
\end{equation}
and
\begin{equation}
\sum_{i=1}^{N} \sum_{j=1}^{N} L^{(0)}_{ij} X_i Z_j (1 -
\delta_{ij}) = \sum_{i=1}^{N} \sum_{j=1}^{N} L^{(0)}_{ij} X_j Z_i (1 - \delta_{ij})~,
\end{equation}
one can write
\begin{equation}
\mathcal{T}_2 = \sum_{i=1}^{N} \sum_{j=1}^{N} L^{(0)}_{ij} \mathcal{R} \left(X_i
  Z_j + X_j Z_i - X_i Z_i -  X_j Z_j\right) (1 - \delta_{ij})~.
\end{equation}
Writing the term in $Z_i Z_j$ in a similar manner as $\mathcal{T}_1$, one
finally obtains
\begin{equation}\label{expr sigmas}
\sigma_s = \sum_{\scriptsize\begin{array}{c}
    i,j = 1\\ i < j \end{array}}^{N} (-L^{(0)}_{ij}) \ I_{ij}~,
\end{equation}
where
$$
I_{ij} = (X_i - X_j)^2
  + (Z_i - Z_j)^2 -  2 \mathcal{R} C_{ij}~, \hspace{5mm} C_{ij} = X_i Z_j + X_j Z_i - X_i Z_i - X_j Z_j~.
$$
This is the relation \eref{sigmas expression} used in \Subsection~\ref{The
  entropy production} to prove that $\sigma_s \geq 0$.

\renewcommand{\theequation}{D.\arabic{equation}}
\setcounter{equation}{0} 
\section*{Appendix D: Equilibrium and Non-Equilibrium States}

In this appendix, we prove the equivalent relations presented in
\Subsection~\ref{Equilibrium
  and non-equilibrium states}. 

To obtain the first one,
$$
\{\mbox{System is at equilibrium}\} \iff \{I_i = 0 \hspace{3mm}
\mbox{and} \hspace{3mm} J_i = 0, \hspace{3mm} \forall i\}~,
$$
let us write the no-current condition ($i\in\{\L, \R, 1, \dots, N\}$):
\begin{eqnarray}
I_i &=& \sum_j L_{ij}^{(0)} \, \frac{\delta \mu_j}{e} + L_{ij}^{(1)}
\, \frac{\delta T_j}{T} = 0~,\label{cond 1 I}\\
J_i &=& \sum_j L_{ij}^{(1)} \, \frac{\delta \mu_j}{e} + L_{ij}^{(2)}
\, \frac{\delta T_j}{T} = 0~.\label{cond 1 J}
\end{eqnarray}
It is convenient to define $X, Y \in \mathbb{R}^{N+2}$ by
\begin{equation}
X_j = \frac{\delta \mu_j}{e} \hspace{5mm} \mbox{and} \hspace{5mm} Y_j
= \frac{\delta T_j}{T}~.
\end{equation}
Then, using the relations $L_{ij}^{(1)} = Q_1 L_{ij}^{(0)}$  and $L_{ij}^{(2)} = Q_2
L_{ij}^{(0)}$, and setting $Q = Q_2 / Q_1 $,  one sees that the conditions \eref{cond 1 I}--\eref{cond 1 J} can be
rewritten as follows:
\begin{eqnarray}\label{XY}
L^{(0)} (X + Q_1  Y) &=& 0~,\label{XY1} \\
L^{(0)} (X + Q  Y)&=& 0~.\label{XY2}
\end{eqnarray}
Since $\mathcal{R} \not = 1$, it follows that $Q_1
\not = Q$. Assume now that all the currents vanish. Then the equations
  \eref{XY1}--\eref{XY2} imply that 
\begin{equation}
L^{(0)} X = 0 \hspace{5mm} \mbox{and} \hspace{5mm} L^{(0)} Y = 0~.
\end{equation}
Since $\sum_j L^{(0)}_{ij} = 0$, the equilibrium
situation, $X_i = \delta \mu/e$ and $Y_i
= \delta T/T$, is a solution.
Observe now that $(-L^{(0)})$ may be interpreted as the generator of an irreducible Markov chain and, consequently, the equation $L^{(0)} V = 0$
has a unique (normalised) solution. Hence
the system is at equilibrium. Finally, the reverse implication is immediate since the equilibrium
situation satisfies the equations \eref{XY1}--\eref{XY2}. 

We next turn to the equivalences in terms of the entropy production:
$$
\left\{\hspace{-1mm}\begin{array}{cc}\mbox{System is at}\\\mbox{equilibrium}\end{array}\hspace{-1mm}\right\} \iff \sigma_s = 0 \hspace{5mm} \mbox{and}
\hspace{5mm} \left\{\hspace{-1mm}\begin{array}{cc}\mbox{System is}\\\mbox{out of
  equilibrium} \end{array}\hspace{-1mm}\right\}\iff \sigma_s > 0~.
$$
Observe that, since $\sigma_s
 \geq 0$, the two above statements are contrapositive
 equivalences. Therefore, it is sufficient to prove the first. For this we use the expression  \eref{expr sigmas} for $\sigma_s$.
 
\vspace{4mm}
\noindent
[$\Rightarrow$] Assume the system is at equilibrium. Then, $X_1 = \dots = X_{N}$ and $Z_{1} = \dots =
 Z_{N}$, and therefore $I_{ij} = 0$ for all $i,j$, from which it
 follows that $\sigma_s = 0$.

\vspace{4mm}
\noindent
[$\Leftarrow$] Conversely, assume that $\sigma_s = 0$. Then, since $(-L^{(0)}_{ij}) < 0$
 and $I_{ij} \geq 0$ for all $i < j$, one has $I_{ij} = 0$ for
 all $i < j$. Therefore, by (\ref{In1a}), one deduces that $X_1 = \dots = X_{N}$ and $Z_{1} = \dots =
 Z_{N}$, \ie the system is at equilibrium. 

\renewcommand{\theequation}{E.\arabic{equation}}
\setcounter{equation}{0} 
\section*{Appendix E: The Matrix $\mathbf{L^{(0)}_{\rm C}}$}

Here we prove that the
matrix $L^{(0)}_{\rm C}$ is real positive definite and consequently
that $L^{(0)}_{\rm C}$ has positive determinant and is therefore invertible. 

Let $W \in \mathbb{R}^{N}$, $W \not = 0$, and 
\begin{equation}
\mathcal{T} \equiv (W, L^{(0)}_{\rm C} W)  = \sum_{i=1}^{N} \sum_{j=1}^{N} L^{(0)}_{ij} W_i W_j~.
\end{equation} 
The idea is to proceed as in the proof of $\sigma_s \geq 0$, but
with the identity
\begin{equation}
L^{(0)}_{ii} = - L^{(0)}_{i\L} - L^{(0)}_{i\R} - \sum_{j=1}^{N}
L^{(0)}_{ij} (1-\delta_{ij})~.
\end{equation}
We obtain
\begin{equation}
\mathcal{T} = -\sum_{\scriptsize\begin{array}{c}
    i,j = 1\\ i < j \end{array}}^{N} L^{(0)}_{ij} (W_i- W_j)^2 -
    \sum_{i=1}^{N} (L^{(0)}_{i\L} + L^{(0)}_{i\R}) W_i^2 > 0~.
\end{equation}
This shows that $L^{(0)}_{\rm C}$ is real positive definite. 

\newpage

Now, let $\lambda_1,
\dots, \lambda_N$ denote the eigenvalues of
$L^{(0)}_{\rm C}$, which might be complex since $L^{(0)}_{\rm C}$ is not symmetric in general. 
\begin{itemize}
\item[(1)] Assume that $\lambda$ is a real eigenvalue of $L^{(0)}_{\rm C}$ and let $v \in
  \mathbb{R}^{N}$ be a normalised associated eigenvector. Then
\begin{equation}
\lambda = \lambda \| v\| = (v, \lambda v) = (v,  L^{(0)}_{\rm C}
v) > 0~,
\end{equation}
where we have used that $L^{(0)}_{\rm C}$ is real positive definite.
\item[(2)] Assume next that $\lambda$ is a complex eigenvalue:
  $\lambda = a + i b$, with $a,b \in \mathbb{R}$ and $b \not = 0$. Then, since $L^{(0)}_{\rm C}$ is a real matrix, both
  $\lambda$ and its complex conjugate $\overline{\lambda}$ are
  eigenvalues of $L^{(0)}_{\rm C}$.
\end{itemize}
In summary, we have shown that the eigenvalues of $L^{(0)}_{\rm C}$ are either
positives or come in pairs $(\lambda, \overline{\lambda})$ with
$\lambda \not = 0$. Therefore
\begin{equation}  
\det(L^{(0)}_{\rm C}) = \prod_i \lambda_i > 0~.
\end{equation}

\renewcommand{\theequation}{F.\arabic{equation}}
\setcounter{equation}{0} 
\section*{Appendix F: The Profiles}

Here we explain the details in the derivation of the profiles
  \eref{Eq2 Profile}--\eref{Eq2 Profileb}. From \eref{Equ MXY},
  one sees that the profiles are
  obtained by solving the generic equation:
\begin{equation}
L^{(0)}_{\rm C} X = - \sum_{\ell = \L, \R}  D_\ell X_\ell~.
\end{equation}
From Appendix~E, the matrix $L^{(0)}_{\rm C}$ can be
    inverted to give
\begin{equation}
X = - \sum_{\ell = \L, \R} [L^{(0)}_{\rm C}]^{-1} D_\ell X_\ell~.
\end{equation}
Recalling that $(D_{\ell})_j = L^{(0)}_{j \ell}$, one has in components:
\begin{equation}
X_i = - \sum_{j = 1}^{N} ([L^{(0)}_{\rm C}]^{-1})_{ij} \left[L^{(0)}_{j \L}
  X_\L + L^{(0)}_{j \R}
  X_\R \right]~.
\end{equation}
Now comes a trick:
\begin{equation}
L^{(0)}_{j \L} = - L^{(0)}_{j \R} - \sum_{k = 1}^{N} L^{(0)}_{j k}~.
\end{equation}
With this identity, one has
\begin{eqnarray}
X_i &=& \sum_{k = 1}^{N} \sum_{j = 1}^{N} ([L^{(0)}_{\rm C}]^{-1})_{ij}
L^{(0)}_{j k} X_\L - \sum_{j = 1}^{N} ([L^{(0)}_{\rm C}]^{-1})_{ij}
L^{(0)}_{j \R} (X_\R - X_\L)\nonumber\\
&=& X_\L - \sum_{j = 1}^{N} ([L^{(0)}_{\rm C}]^{-1})_{ij}
L^{(0)}_{j \R} (X_\R - X_\L)~.
\end{eqnarray}
Observe that $(L^{(0)}_{C})_{ij} =
  L^{(0)}_{ij} = e^2 C(0)/h \ \Gamma_{ij}$ and
  $L^{(0)}_{j\R} = - e^2 C(0)/h \ t_{j\R}$. Hence, the constant $e^2 C(0)/h$ involved in 
$([L^{(0)}_{\rm C}]^{-1})_{ij}$ and in $L^{(0)}_{j\R}$ simplifies,
  and we are left with profiles independent of the distribution $f$
  describing the reservoirs. The temperature and chemical potential
  profiles are obtained by setting $X_i = \delta T_i / T$, with $T_i =
T + \delta T_i$, and $X_i = \delta \mu_i / e$, with $\mu_i =
\mu + \delta \mu_i$, respectively.

\renewcommand{\theequation}{G.\arabic{equation}}
\setcounter{equation}{0} 
\section*{Appendix G: An Expression for $\mathbf{A_i}$}\label{Appendix Reminder}
In this appendix, we derive the alternative expression \eref{Prop B}
for the coefficient $A_i$. From \eref{Eq2 Profileb}, one has
\begin{equation}
A_i = \sum_{j = 1}^{N} (\Gamma_{\rm C}^{-1})_{ij} \ t_{j \R}~.
\end{equation}
Now, we write
\begin{equation}\label{equ gammacinv}
(\Gamma_{\rm C}^{-1})_{ij} = \frac{1}{\det\left(\Gamma_{\rm C}\right)} (-1)^{i+j} \det\left(\Gamma_{\rm C}(j,i)\right)~,
\end{equation}
so that
\begin{equation}
A_i = \frac{1}{\det\left(\Gamma_{\rm C}\right)} \sum_{j = 1}^{N} (-1)^{i+j}
\det\left(\Gamma_{\rm C}(j,i)\right) t_{j \R}~.
\end{equation}
It remains to expand $\det\left(\Gamma_{\rm C}\right)$. One has
\begin{equation}
\det\left(\Gamma_{\rm C}\right) =  \sum_{j = 1}^{N} (-1)^{i+j} \det\left(\Gamma_{\rm C}(j,i)\right) \Gamma_{ji}~.
\end{equation}
Using the relation
\begin{equation}
\Gamma_{ji} = - \Gamma_{j \L} - \Gamma_{j \R} - \sum_{\scriptsize\begin{array}{c}
    k = 1\\ k \not
= i\end{array}}^{N} \Gamma_{j k}~,
\end{equation}
one obtains
\begin{equation}
\det\left(\Gamma_{\rm C}\right) = \sum_{j = 1}^{N} (-1)^{i+j}
\det\left(\Gamma_{\rm C}(j,i)\right) [t_{j \L} + t_{j \R}] - R_i~,
\end{equation}
where
\begin{equation}
R_i = \sum_{j = 1}^{N} \sum_{\scriptsize\begin{array}{c}
    k = 1\\ k \not
= i\end{array}}^{N} (-1)^{i+j}
\det\left(\Gamma_{\rm C}(j,i)\right) \Gamma_{j k}~.
\end{equation}
Finally, using \eref{equ gammacinv}, one obtains
\begin{equation}
R_i = \det\left(\Gamma_{\rm C}\right) \sum_{\scriptsize\begin{array}{c}
    k = 1\\ k \not
= i\end{array}}^{N} \sum_{j = 1}^{N}  \left(\Gamma_{\rm
    C}^{-1}\right)_{ij}  \Gamma_{j k} = \det\left(\Gamma_{\rm
    C}\right) \sum_{\scriptsize\begin{array}{c}
    k = 1\\ k \not
= i\end{array}}^{N} \delta_{ik} = 0~.
\end{equation}

\renewcommand{\theequation}{H.\arabic{equation}}
\setcounter{equation}{0} 
\section*{Appendix H: The Global Scattering Matrix $\mathbf{S_N}$}\label{General case S}

In this appendix, we explain how to build the scattering matrix $S$ of the
system made of $N$ quantum dots from the $N$ individual scattering matrices
$S^{(1)},\dots,S^{(N)}$, where $S^{(k)}$ is the $(M + M_k + M) \times (M
+ M_k + M)$ unitary scattering matrix associated with the $k$-th quantum dot. Here,
we have set $M_\L = M_\R = M$.

\begin{figure}[h!]
\begin{center}
\psfig{file=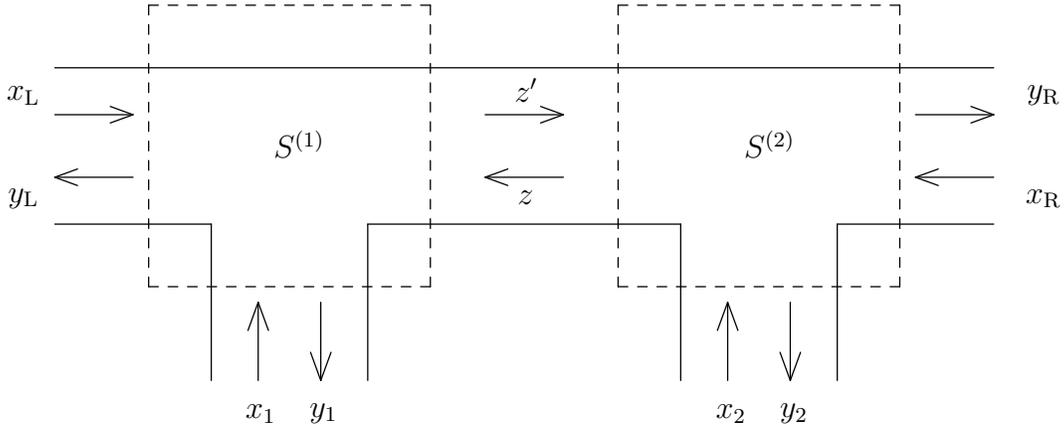,width=12.5cm}
\vspace{15mm}
\caption{Combination of $S^{(1)}$ and $S^{(2)}$ into the composite S-matrix
  $S_2$.}\label{Figure S}
\end{center}
\end{figure}

Let us start with the case $N=2$. Let $S^{(1)}$ and $S^{(2)}$ denote the scattering matrices of the first
and second quantum dot, respectively, and let $S_2$ denote the
corresponding scattering matrix for the system made of two
quantum dots  (see \Fig~\ref{Figure S}).

From \Fig~\ref{Figure S}, one easily sees that these matrices are related as follows:
\begin{equation}\label{N equal to 1}
S^{(1)} \matrice{x_\L \\ x_1 \\ z } =
\matrice{y_\L \\ y_1 \\ z' }~, \hspace{5mm} S^{(2)} \matrice{z' \\ x_2 \\ x_\R } =
\matrice{z \\ y_2 \\ y_\R }~, \hspace{5mm} S_{2}
\matrice{x_\L \\ x_1 \\ x_2 \\ x_\R } =
\matrice{y_\L \\ y_1 \\ y_2 \\ y_\R }~.
\end{equation}
Introducing 
\begin{equation}
r^{(1)} = \matrice{ S^{(1)}_{11} & S^{(1)}_{12} \\ S^{(1)}_{21}
  & S^{(1)}_{22}}~, r'^{(1)} = S^{(1)}_{33}~, t^{(1)} =
  \matrice{ S^{(1)}_{31} & S^{(1)}_{32} }~,  t'^{(1)} =
  \matrice{ S^{(1)}_{13} \\ S^{(1)}_{23} }
\end{equation}
and
\begin{equation}
x_{\L 1} = \matrice{ x_\L \\ x_1 }~,
y_{\L 1} = \matrice{ y_\L \\ y_1 }
\end{equation}
one can rewrite the first relation in \eref{N equal to 1} as
follows:
\begin{equation}\label{Equ S1}
\matrice{ r^{(1)} & t'^{(1)} \\ t^{(1)} & r'^{(1)}}
\matrice{ x_{\L 1} \\ z } = \matrice{ y_{\L 1} \\ z' }~.
\end{equation}
Similarly, setting
\begin{equation}
r^{(2)} = S^{(2)}_{11}~,   r'^{(2)} = \matrice{ S^{(2)}_{22} & S^{(2)}_{23} \\ S^{(2)}_{32}
  & S^{(2)}_{33}}~, t^{(2)} =
  \matrice{ S^{(2)}_{21} \\ S^{(2)}_{31} }~,  t'^{(2)} =
  \matrice{ S^{(2)}_{12} & S^{(2)}_{13} }
\end{equation}
and
\begin{equation}
x_{2 \R} = \matrice{ x_2 \\ x_\R }~,
y_{2 \R} = \matrice{ y_2 \\ y_\R }
\end{equation}
one has for the second relation in \eref{N equal to 1}:
\begin{equation}\label{Equ S2}
\matrice{ r^{(2)} & t'^{(2)} \\ t^{(2)} & r'^{(2)}}
\matrice{ z' \\  x_{2 \R}} = \matrice{ z \\  y_{2 \R} }~.
\end{equation}
Finally, we write the third relation in \eref{N equal to 1} as
follows:
\begin{equation}\label{Equ S12}
\matrice{ r_{2} & t'_{2} \\ t_{2} & r'_{2}}
\matrice{ x_{\L 1} \\  x_{2 \R}} = \matrice{
  y_{\L 1} \\  y_{2 \R} }~.
\end{equation}
Solving the equations~\eref{Equ S1} and \eref{Equ S2} for $y_{\L 1}$ and
$y_{2 \R}$, and recalling \eref{Equ S12}, one obtains
\begin{eqnarray}
r_{2} &=& r^{(1)} + t'^{(1)} [1 - r^{(2)} r'^{(1)}]^{-1} r^{(2)}
t^{(1)}~,\\
r'_{2} &=& r'^{(2)} + t^{(2)} [1 - r'^{(1)} r^{(2)}]^{-1} r'^{(1)}
t'^{(2)}~,\\
t_{2} &=& t^{(2)} [1 - r'^{(1)} r^{(2)}]^{-1} t^{(1)}~,\\
t'_{2} &=& t'^{(1)} [1 - r^{(2)} r'^{(1)}]^{-1} t'^{(2)}~.
\end{eqnarray}
Let us now generalise these relations to the $N > 2$ quantum dots situation. For
this we denote by $S^{(1)},\dots,S^{(N)}$ the $N$ individual
scattering matrices and by $S_{N}$ the composite S-matrix associated to
$N$ quantum dots. We write
\begin{equation}
S_{N} \matrice{x_\L \\ x_1 \\ : \\ x_N \\x_\R
} = \matrice{y_\L \\ y_1 \\ : \\ y_N \\ y_\R
}~, \hspace{10mm} S_{N} = \matrice{ r_{N} & t'_{N} \\ t_{N} & r'_{N}}~.
\end{equation} 
Then
\begin{eqnarray}
r_{N} &=& r^{(12\dots N-1)} + t'^{(12\dots N-1)} [1 - r^{(N)} r'^{(12\dots N-1)}]^{-1} r^{(N)}
t^{(12\dots N-1)}~,\label{Eq rN}\\
r'_{N} &=& r'^{(N)} + t^{(N)} [1 - r'^{(12\dots N-1)} r^{(N)}]^{-1} r'^{(12\dots N-1)}
t'^{(N)}~,\\
t_{N} &=& t^{(N)} [1 - r'^{(12\dots N-1)} r^{(N)}]^{-1} t^{(12\dots N-1)}~,\\
t'_{N} &=& t'^{(12\dots N-1)} [1 - r^{(N)} r'^{(12\dots N-1)}]^{-1} t'^{(N)}~,\label{Eq tpN}
\end{eqnarray}
where
\begin{equation}
r^{(N)} = S^{(N)}_{11}~, \; r'^{(N)} = \matrice{ S^{(N)}_{22} & S^{(N)}_{23} \\ S^{(N)}_{32}
  & S^{(N)}_{33}}~, \; t^{(N)} =
  \matrice{ S^{(N)}_{21} \\ S^{(N)}_{31} }~,  \;t'^{(N)} =
  \matrice{ S^{(N)}_{12} & S^{(N)}_{13} }~.
\end{equation}
and
\begin{equation}
r^{(12\dots N-1)} = \matrice{ (S_{N-1})_{11} & \dots &
  (S_{N-1})_{1N} \\ : & & : \\
  (S_{N-1})_{N1} & \dots & (S_{N-1})_{NN}}~, \; r'^{(12\dots N-1)} =
  (S_{N-1})_{N+1 \ N+1}~, 
\end{equation}
\begin{equation}
t^{(12\dots N-1)} =
  \matrice{ (S_{N-1})_{N+1 \ 1} & \dots &  (S_{N-1})_{N+1 \ N}}~, \; t'^{(12\dots N-1)} =
  \matrice{ (S_{N-1})_{1 \ N+1} \\ : \\ (S_{N-1})_{N \ N+1} }~.
\end{equation}
Therefore, by working recursively, one can build the global scattering
matrix $S = S_{N}$ from the local scattering matrices
$S^{(1)},\dots,S^{(N)}$. Note that, by construction, if
$S^{(1)},\dots,S^{(N)}$ are unitary, then $S$ will also be unitary.\\ \\
\noindent
\textit{Remark.} If one considers that the terminals have only one channel: $M_i = 1$ for all
$i$, then $r^{(N)}$ and $r'^{(12\dots N)}$ become complex numbers and
we can introduce the following complex number:
\begin{equation}
c_N = [1 - r^{(N)} r'^{(12\dots N-1)}]^{-1}~.
\end{equation}
Then one can rewrite the equations~\eref{Eq rN}--\eref{Eq tpN} as follows:
\begin{eqnarray}
r_{N} &=& r^{(12\dots N-1)} + c_N \ r^{(N)} \ t'^{(12\dots N-1)} t^{(12\dots N-1)}~,\label{Eq rN2}\\
r'_{N} &=& r'^{(N)} + c_N \ r'^{(12\dots N-1)} \ t^{(N)} t'^{(N)}~,\\
t_{N} &=& c_N \ t^{(N)} t^{(12\dots N-1)}~,\\
t'_{N} &=& c_N \ t'^{(12\dots N-1)} t'^{(N)}~.\label{Eq tpN2}
\end{eqnarray}
These are the equations that are used in the numerical simulations in \Section~\ref{Numerical analysis}.


\bibliographystyle{unsrt}
\bibliography{refs}

\end{document}